\documentclass[1p]{elsarticle}
\usepackage{amssymb}
\usepackage{amsmath}
\usepackage{amsfonts}
\usepackage{graphicx}
\usepackage{epstopdf}
\usepackage{epsfig}
\newcommand{\B}[1]{\mbox{\boldmath${#1}$\unboldmath}}

\usepackage{soul}
\usepackage{color}

\renewcommand{\a}[1]{\mathbf{a}_{#1}}
\renewcommand{\b}[1]{\mathbf{b}_{#1}}
\newcommand{\e}[1]{\mathbf{e}_{#1}}
\newcommand{\bq}[1]{\tilde{\mathbf{b}}_{#1}}
\newcommand{\R}{\mathbf{R}}
\renewcommand{\r}{\mathbf{r}}
\newcommand{\K}{\mathbf{G}}
\renewcommand{\k}{\mathbf{k}}
\newcommand{\Q}{\mathbf{Q}}
\newcommand{\q}{\mathbf{q}}
\newcommand{\ee}{ \mathrm{e}}
\newcommand{\ii}{ \mathrm{i}}

\usepackage[normalem]{ulem}
\definecolor{JBblue}{RGB}{0, 0, 255}

\begin{document}


\newcommand{\fracc}[2]{\frac{\displaystyle #1}{\displaystyle #2}} 

\begin{frontmatter}
\title{Frequency-momentum representation of moving breathers in a two dimensional hexagonal lattice}%
\author[lu]{J\={a}nis Baj\={a}rs}
\address[lu]{Faculty of Physics, Mathematics and Optometry,University of Latvia, Jelgavas Street 3, Riga, LV-1004, Latvia}
\author[gfnl]{Juan F. R. Archilla\corref{cor1}}
 \address[gfnl]{Group of Nonlinear Physics, Universidad de Sevilla, ETSII, Avda Reina Mercedes s/n, 41012-Sevilla, Spain}
 \ead{archilla@us.es}
\cortext[cor1]{Corresponding author}

\begin{abstract}
We study nonlinear excitations propagating in a hexagonal layer which is a model for the cation layer of silicates. We consider their properties in the frequency-momentum or $\omega-k$ representation, extending the theory on pterobreathers in their moving frame for the first time to two dimensions. It can also be easily extended to three dimensions. Exact traveling waves in the $\omega-k$ representation are within {\em resonant} planes, each plane corresponding in the moving frame to a single frequency. These frequencies are  integer multiples of a frequency called the fundamental frequency.  A breather is within a resonant plane called the breather plane and has a single frequency in the moving frame. The intersection of the resonant planes with the phonon surfaces produce co-traveling wings with a small set of frequencies. The traveling waves obtained by perturbing the system consist of a breather and a soliton traveling together and are quasi-exact. These traveling waves can be used as seeds to obtain exact traveling waves, also formed by a breather and a soliton. The wings do exist but they are usually very small.
\end{abstract}
\end{frontmatter}



\section{Introduction}
\label{sec:introduction}
The study of nonlinear excitations as breathers in a hexagonal lattice is motivated by the experimental evidence of the existence of nonlinear excitations, also known as intrinsic localized modes (ILM), that transport energy and momentum in a quasi-one-dimensional form along the chains of the cation layer of muscovite mica and other layered silicates\,\cite{russell-experiment2007,russell-archilla2017,russell-archilla2019,russell-archilla2021rrl}. These lattice excitations were called {\em quodons} making reference to the quasi-one-dimensional propagation and the particle-like behaviour identified in the fossil dark tracks in muscovite\,\cite{russell-collins95a,russell-collins95b}. It is worth noting that muscovite has been demonstrated able to record the passage of swift particles as positrons, protons and antimuons\,\cite{russell67a,russell67b}. The reader is referred to three complete reviews\,\cite{russell-tracks-quodons2015article,russell-crystal-quodons2015article,russell-archilla2021springer}.

There have been several attempts to model these nonlinear excitations in one dimension as kinks or crowdions\,\cite{archilla_ujp,archilla-kosevich-springer2015article,archilla-kosevich-pre2015,archilla-zolotaryuk2018} and breathers\,\cite{archilla2019}.
These models use a substrate potential and interaction potential that have been deduced from physical properties and empirical potentials, and therefore are able to provide physical values of the magnitudes involved as, for example, energies, $\simeq 27$\,eV for crowdions and $\simeq$ 0.3\,eV for breathers. The latter article \,\cite{archilla2019} also develops the theory of exact breather solutions, which are generally coupled to an extended plane wave called wing and are therefore  called {\em pterobreathers}, ``ptero'' meaning ``wing'' in Greek. However, for some specific velocities the wings disappear and the exact pterobreathers become exact breathers. Their stability can be determined by a variant of the Floquet method. The same article also characterized exact pterobreathers and breathers by their unique frequency in the moving frame. These models have the obvious limitation of being one-dimensional but this limitation also allows for easier mathematical and numerical insight into the physical and mathematical properties.

Other model, inspired by the same phenomena, was proposed\,\cite{marin-eilbeck-russell-2Dhexa1998,bajars-physicad2015} using a more generic model in a two dimensional (2D) hexagonal lattice also with a substrate potential. They obtained that kinks or crowdions can propagate in a quasi-dimensional form\,\cite{bajars-quodons2015article}  and also breathers\,\cite{marin-eilbeck-russell-2Dhexa1998,bajars-physicad2015}, depending on the specifics of the potential. Breathers are also scattered by other breathers and can migrate to other close-packed chains\,\cite{bajars2021}.

In this paper  we extend the theory developed in \cite{archilla2019} to
two dimensions and apply it to the model in Ref.\,\cite{bajars-physicad2015} in order to describe moving breathers in the moving frame in a hexagonal lattice with a substrate potential, to obtain their frequency-momentum representation, their frequencies in the moving frame and the existence or absence of wings. This is  fundamental  to be able to interpret possible signatures of localized nonlinear waves in physical spectra of real crystals\,\cite{manley2019}.

The paper is written illustrating the theory together with particular analytical and numerical results. The mathematical model is briefly described in Sect.\,\ref{sec:model}. The on-site potential used to describe the direct hexagonal lattice and the phonon bands to describe the reciprocal and momentum lattices are presented  in Sect.\,\ref{sec:hexagonal}. The phonon frequencies and polarization are obtained in Sect.\,\ref{sec:phonons}. The theory of exact traveling waves in 1D and 2D is developed in Sect.\,\ref{sec:basictheory}. In Sect.\,\ref{sec:quasiexact} quasi-exact solutions are used to obtain important characteristic parameter values for the exact propagating modes. The procedure to obtain exact traveling wave solutions is explained in Sec.\,\ref{sec:newton} and the corresponding exact solutions are presented in Sect.\,\ref{sec:results}. The paper is concluded with the Conclusions and an appendix containing some details of the reciprocal basis.

\section{Model}
\label{sec:model}
We consider a system of $N$ particles with mass $m=1$ with their position described by coordinates $(x,y)$ in a Cartesian system of reference and $(u_x, u_y)$ with respect to their equilibrium positions. Their kinetic energy is given by $K=\tfrac{1}{2}(\dot x^2+ \dot y^2)=\tfrac{1}{2}(\dot u_x^2+ \dot u_y^2)$, the dot indicating the derivative with respect to time.

The particles are in their  equilibrium position in a 2D hexagonal lattice corresponding to the minima of the on-site potential given in the dimensionless form by:
\begin{align}\label{eq:OnSiteFunc}
\begin{split}
  U(x,y) = &\frac{2}{3} \biggl(1-\frac{1}{3}
  \biggl( \cos{\biggl(2\pi\frac{2}{\sqrt{3}}y \biggr)} \biggr. \biggr. \\
  & \qquad + \biggl. \biggl. \cos{ \biggl( 2\pi (x- \frac{1}{\sqrt{3}}y)\biggr)} +
  \cos{ \biggl( 2\pi (x+ \frac{1}{\sqrt{3}}y)\biggr) }
    \biggr) \biggr),
\end{split}
\end{align}
where the unit distance has been chosen $a=1$ and $0\leq U(x,y)\leq 1$.

The interaction between particles is given by the Lennard-Jones potential:
\begin{equation}\label{eq:LJpot}
 V_{LJ}(r) = \epsilon \left( \left( \frac{1}{r} \right)^{12}
- 2 \left( \frac{1}{r} \right)^{6} \right),
\end{equation}
being $\epsilon$ a measure of the relative strength of the interaction potential with respect to the on-site potential.

Without loss of generality we consider $\epsilon=0.05$. For short and long-range interactions we consider smooth cut-off of the Lennard-Jones potential in Eq.~\eqref{eq:LJpot} with a cut-off radius $r_c$ described in detail in \cite{bajars-physicad2015}. In this paper we consider $r_c=3a$. Derived Hamiltonian equations of dynamics are integrated in time with the second order time reversible symplectic Verlet method. In the following, all numerical examples are performed with time step $\tau = 0.04$ and periodic boundary conditions. The dimensions of the lattice are $N_1=64$ and $N_2=32$, i.e., $N=N_1N_2$, unless stated otherwise.

To produce quasi-exact discrete breather solutions, see Sec.~\ref{sec:quasiexact}, we excite three neighboring particle velocities with the pattern:
\begin{equation}\label{eq:pattern}
v_0 = \gamma (-1; 2; -1)^T,
\end{equation}
where $\gamma>0$. As noted in\,\cite{bajars2021} larger values of $\gamma$ produce faster moving quasi-exact breathers with larger particle displacements in the direction of propagation. See also Table \ref{table:exactbreathers}.

\begin{figure}[t]
\begin{center}
\includegraphics[width=0.7\textwidth]{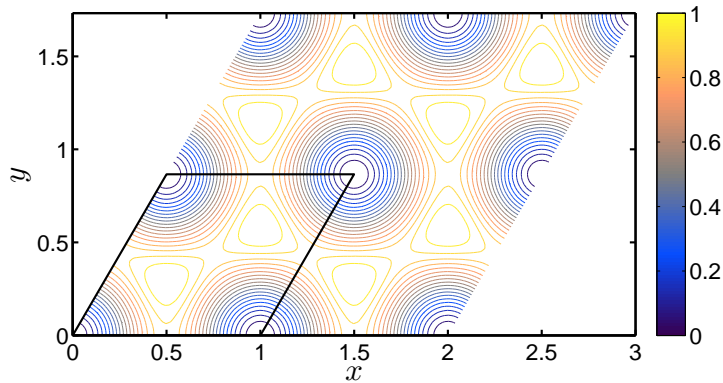}
\end{center}
\caption{Contour plot of the on-site potential $U(x,y)$ of Eq.\,(\ref{eq:OnSiteFunc}) in 4 primitive direct cells and the contours of one of them.  The bottom and left sides correspond to the vectors of the direct primitive basis. Blue colors are lower energies, being the minima at the equilibrium particle positions. The red-yellow colors are higher energies as specified by the color bar and correspond to potential barrier between sites. The associated face-centered rectangular lattice can be visualized above abscissas 1 and 2.  Note that the breather direction along a close-packed line as the horizontal axis has no perpendicular close-packed plane. See Fig.\,\ref{fig_Udirect3D} for a view of $U$ in a single primitive cell.
}
\label{fig_Udirect}
\end{figure}

\section{The 2D hexagonal lattice}
\label{sec:hexagonal}

\begin{figure}[ht]
\begin{center}
\includegraphics[width=0.7\textwidth]{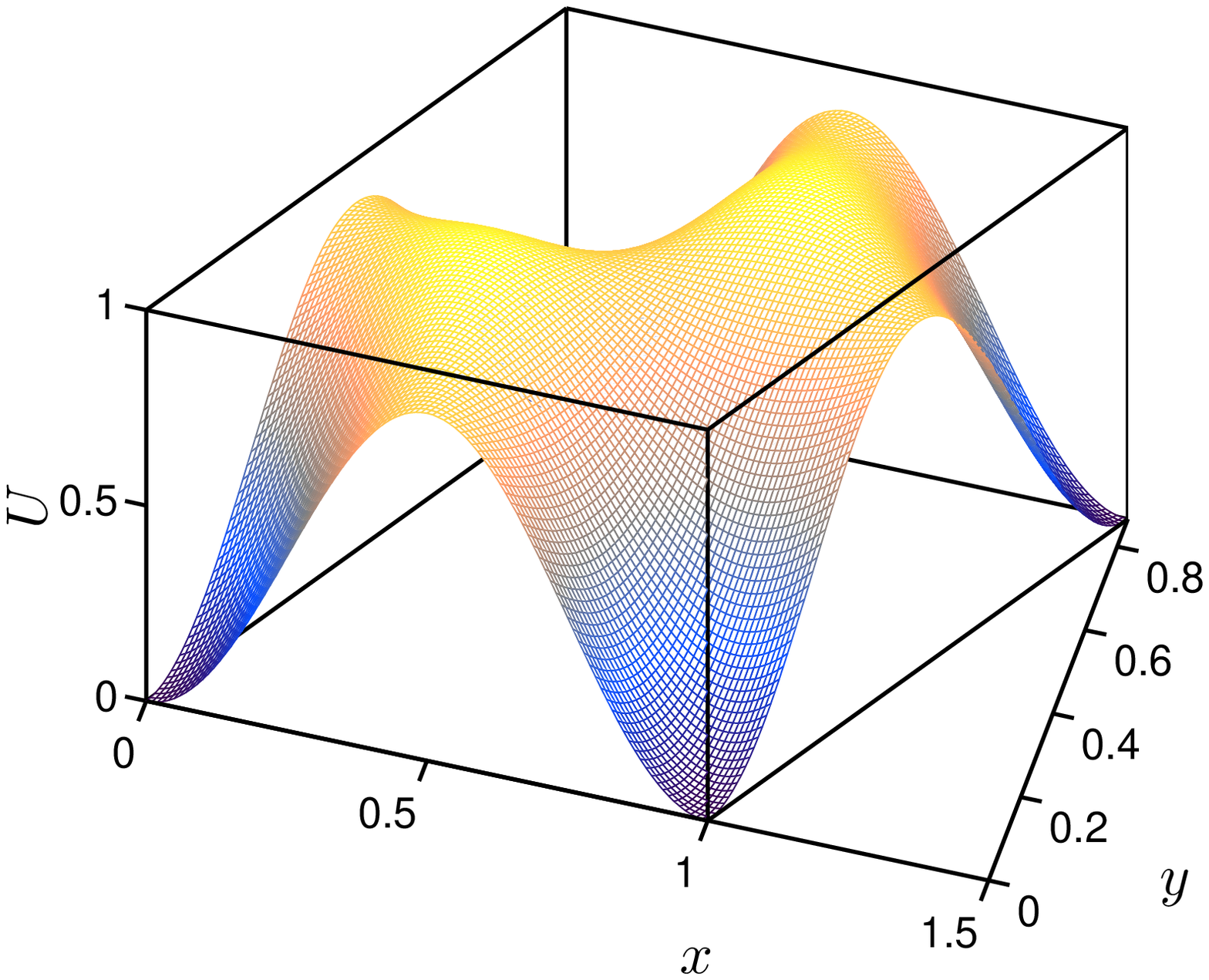}
\end{center}
\caption{Surface $U(x,y)$ of Eq.\,(\ref{eq:OnSiteFunc}) in a single primitive direct cell. Note the difference in the paths between nearest neighbors at the corner of the primitive cell or along the short diagonal, which are equal, and along the large diagonal.
}
\label{fig_Udirect3D}
\end{figure}

For understanding the frequency-momentum representation it is convenient to review the properties of the hexagonal lattice. In this section we  perform that revision while at the same time  we present important properties of the system, such as  its potential energy and linear spectrum.

\subsection{The direct 2D hexagonal lattice}
\label{subsec:directhexagonal}
For a 2D hexagonal lattice of unit distance $a=1$, the direct vectors that generate the lattice are (see \ref{App:reciprocal} and Ref.\,\cite{ashcroftmermin}):
\begin{equation}
\a1=\e1; \quad \a2= \frac{1}{2}\e1 +\frac{\sqrt{3}}{2}\e2\, ,
\end{equation}
with $\e1=[1,0]$ and $\e2=[0,1]$, the Cartesian unit vectors. Any lattice point can be obtained as a point of the Bravais direct lattice:
\begin{equation}
{\R }=n_1 \a1+n_2\a2,
\label{eq:directBravais}
\end{equation}
with $n_1, n_2$ integers.

Many physical properties are described by a function with the periodicity of the direct Bravais lattice, for example, the on-site potential in Eq.\,(\ref{eq:OnSiteFunc}) as can be seen in Fig.~\ref{fig_Udirect}. The minima of the potential correspond to the equilibrium position of the particles and form the hexagonal Bravais lattice. In addition, the form of the potential within a primitive cell can be seen  in Fig.\,\ref{fig_Udirect3D}.

The hexagonal lattice can also be seen as a face centered rectangular lattice with sides $1$ and $\sqrt{3}$, which also corresponds to two primitive simple rectangular Bravais lattices displaced half a diagonal vector equal to $\a2$.  Convenient labels in this description are integers $l$ and $m$ such that the lattice points have the coordinates:
\begin{equation}
x(l,m)=\frac{1}{2} l;\quad y(l,m)=\frac{\sqrt{3}}{2}m\, .
\label{eq:indiceslm}
\end{equation}
In this description $l+m$ is even. When $l$ and $m$ are both odd, they describe the lattice points of a rectangular lattice, and when they are both even, they describe the displaced simple rectangular lattice. This description is used often to write the dynamical equations since the visualization is easier\,\cite{ikeda-doi2007,bajars-physicad2015}. It also provides a convenient matrix form with indexes $l,m$, which corresponds to the lattice with a shift of alternate rows 0.5 to the left.  However, it is not a Bravais lattice and therefore it is not convenient for the description in the momentum or reciprocal space.  Both sets of indexes are related by $x(l,m)=\tfrac{1}{2}l=n_1 +\tfrac{1}{2}n_2$ and $y(l,m)=\tfrac{\sqrt{3}}{2} m =\tfrac{\sqrt{3}}{2} n_2$, that is: $l=2 n_1+n_2$ and $m=n_2$, taking into account Born-von Karman periodic conditions.

\begin{figure}[ht]
\begin{center}
\includegraphics[width=0.9\textwidth]{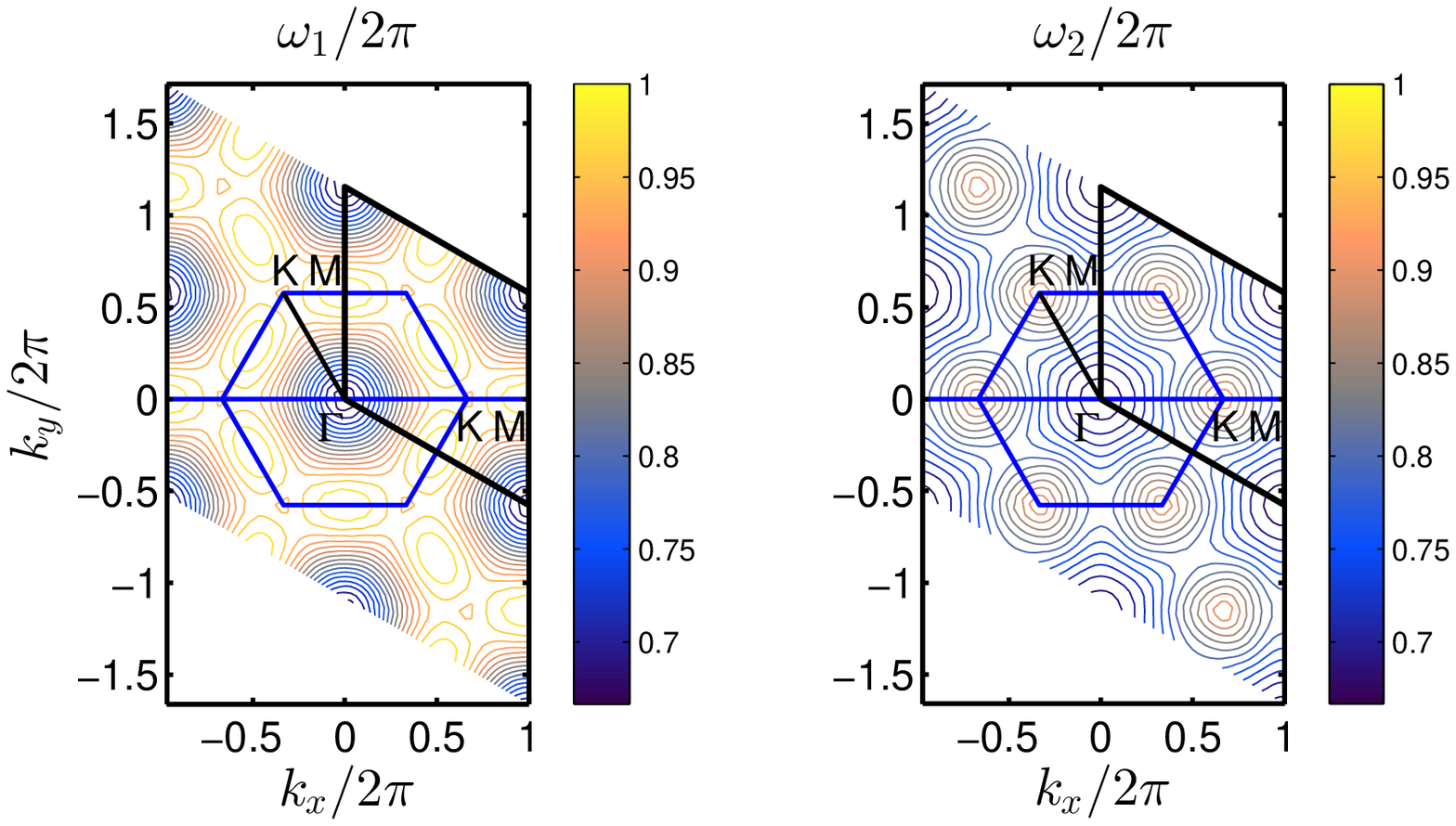}
\end{center}
\caption{Contour plot of the upper and lower phonon bands $\omega_1(k_x,k_y)$ (left) and $\omega_2$ (right)  in 4 primitive reciprocal cells and the contour of one of them. The left and bottom sides correspond to the vectors of the reciprocal basis. Also marked are the horizontal axis, the Brillouin zone and the symmetry points $\Gamma$, M, K, showing the equivalence of two paths $\Gamma$-K-M. At points K the two phonon bands touch each other as can be seen in the $\omega-k$ plot along $\Gamma$-M-$\Gamma$ shown in Fig.~\ref{fig_BrillouinGMKG}.
 }
\label{fig_phononf1f2}
\end{figure}
\subsection{The reciprocal hexagonal lattice}
\label{subsec:reciprocal}
The corresponding reciprocal lattice basis of the direct Bravais lattice in Eq.\,(\ref{eq:directBravais}) is given by:
\begin{equation}
\b1=2\pi [1, -\frac{1}{\sqrt{3}}] ; \quad \b2=2\pi [0, \frac{2}{\sqrt{3}}]\, ,
\end{equation}
with the properties that $|\b1|=|\b2|=2\pi \tfrac{2}{\sqrt{3}}$, and $\a{i}\cdot\b{j}=2\pi\delta_{ij}$.

The (angular) reciprocal  space of Eq.\,(\ref{eq:directBravais}) is the Bravais lattice expanded by $\{\b1,\b2\}$, that is, $\K=m_1\b1+m_2\b3$, with $m_1$ and $m_2$ integers, with the property that $\K\cdot\R=2\pi (m_1 n_1+ m_2 n_2)=2\pi n $, with $n$ an integer.
We can also consider the reciprocal vector $\Q=\K/2\pi$, which has the same meaning as $\K$ but in linear (m$^{-1}$) instead of angular (rad$\cdot$m$^{-1}$) units, similarly to the frequency $f=\omega/2\pi$ and the angular frequency $\omega$.
The appropriate basis for $\Q$ are the vectors $\bq{1}=\b{1}/2\pi=[1,-1/\sqrt{3}]$ and  $\bq{2}=\b{2}/2\pi=[0,2/\sqrt{3}]$ with $\a{i}\cdot\bq{j}=\delta_{ij}$. In this basis,  $\Q=m_1\bq1+m_2 \bq2$ with the same values  $m_1$ and $m_2$ of $\bf G$. $\bf G$ and $\omega$ are sometimes called the physics definition and $\Q$ and $f$ the crystallographic definition of the reciprocal vectors and frequency, respectively\,\cite{vainshtein2010}.
Then:
\begin{equation}
\ee^{ \ii \K\cdot \R }=\ee^{ \ii 2\pi \Q\cdot \R }=1\,.
 \end{equation}

Therefore, any property with the periodicity of the direct Bravais lattice can be described as a sum of exponentials $\exp( \ii \K\cdot \mathbf{r})$.
For example, the on-site potential in Eq.\,(\ref{eq:OnSiteFunc})  is given by:
\begin{equation}
  U(x,y) = \frac{2}{3} \biggl(1-\frac{1}{3}
( \cos(\b2 \cdot \r )  + \cos( \b1 \cdot \r) + \cos( [\b1+\b2]\cdot \r ))\biggr),
\label{eq:OnSiteRecip}
\end{equation}
being the simplest form of a potential in the hexagonal lattice consistent with the symmetries of the lattice. The reciprocal vector $\b1+\b2=2\pi[1,1/\sqrt{3}]$ is symmetric to $\b1$ with respect to the horizontal lattice and it is necessary for the hexagonal symmetry of the potential.

\subsection{The supercell lattice and the wavevector or momentum space}
\label{subsec:momentumspace}
However, as phonons,  breathers or kink solutions do not have the symmetry of the lattice (although the whole set of them has it) and we have Born-von Karman periodic conditions, we have to consider the Bravais direct lattice expanded by $N_1\a1$ and $N_2\a2$, where $N_1$ and $N_2$ are the dimensions of the lattice. It is called the supercell lattice as its primitive cell is usually the computational lattice. The corresponding reciprocal space is the momentum or wavevector space and it is expanded by the basis vectors $\{\b1/N_1, \b2/N_2\}$, or in other form they correspond to:
\begin{equation}
\k=q_1\b1+q_2\b2 \quad \mathrm{with}\quad q_1=0,\frac{1}{N_1},\dots,\frac{N_1-1}{N_1},\quad q_2=0,\frac{1}{N_2},\dots,\frac{N_2-1}{N_2} \, .
\label{eq:momentumspace}
\end{equation}
Any wavevector $\k$ is equivalent to $\k+\K$, with $\K$  a vector in the reciprocal lattice.  The primitive cell of the wavevector space is visualized by the phonon modes obtained in Sect.\,\ref{sec:phonons} as can be seen in Fig.\,\ref{fig_phononf1f2}.  By translations of vectors $\K$, it is often constructed the first Brillouin zone, that is, a zone in momentum space which is closer to a given point than to any other and it is also visualized in the same figure. Important points and paths are often used to visualize the phonon bands as also depicted in the same figure. We can also define the crystallographic wavevectors
$\q=\k/(2\pi)$ to get rid of the factor $2\pi$, then
$\q=q_1\bq{1}+q_2\bq{2}=q_x\e1+q_y\e2$. Note that $q_x=k_x/2\pi$ and $q_y=k_y/2\pi$, but $q_1$ and $q_2$ are the same in both representations. As they have the reciprocal basis vectors as unit of length they are the reduced wave vectors\,\cite{dove1993}.

\begin{figure}[t]
\begin{center}
\includegraphics[width=0.9\textwidth]{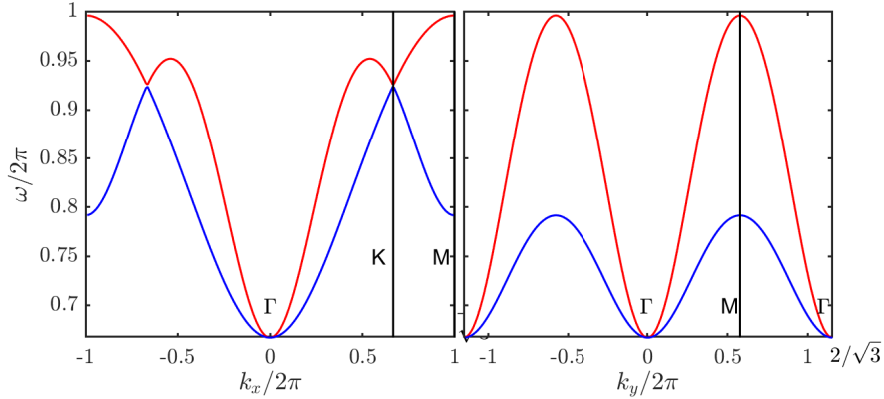}
\end{center}
\caption{Frequency as a function of the wavenumber along the path $\Gamma$-K-M-$\Gamma$ within and at the border of the Brilloin zone as shown in Fig.\,\ref{fig_phononf1f2}. Momenta within the $\Gamma$-K-M  correspond to waves that propagate along the $x$ direction with wavefronts perpendicular to it as the breathers found in this work.  Momenta within $\Gamma$-M correspond to planar waves that propagate parallel to $\b1$ or $\b2$ and perpendicular to direct planes parallel to $\a1$ or $\a2$. Polarization: $\Gamma$-K upper curves have polarization $u_y=0$ and lower ones $u_x=0$, K-M and $\Gamma$-M upper curves have $u_x=0$ and lower ones $u_y$=0.
 }
\label{fig_BrillouinGMKG}
\end{figure}

Any wavevector $\k $ represents a set of planes\footnote{We keep the word {\em planes} for 3D although in 2D they degenerate to  straight lines} perpendicular to $\k$ separated by a distance $2\pi/|\k|=1/|\q|$ or a plane wave with plane wavefronts separated by the same distance.  Therefore, it is better seen as defining a plane than as a direction.  As $\k +\K$  represents the same set of planes, it is enough to consider a primitive reciprocal lattice with non equivalent vectors $\k$, either within the primitive reciprocal lattice, the Brillouin zone or any other construction through translations by vectors $\K$.

\begin{figure}[t]
\begin{center}
\includegraphics[width=0.49\textwidth]{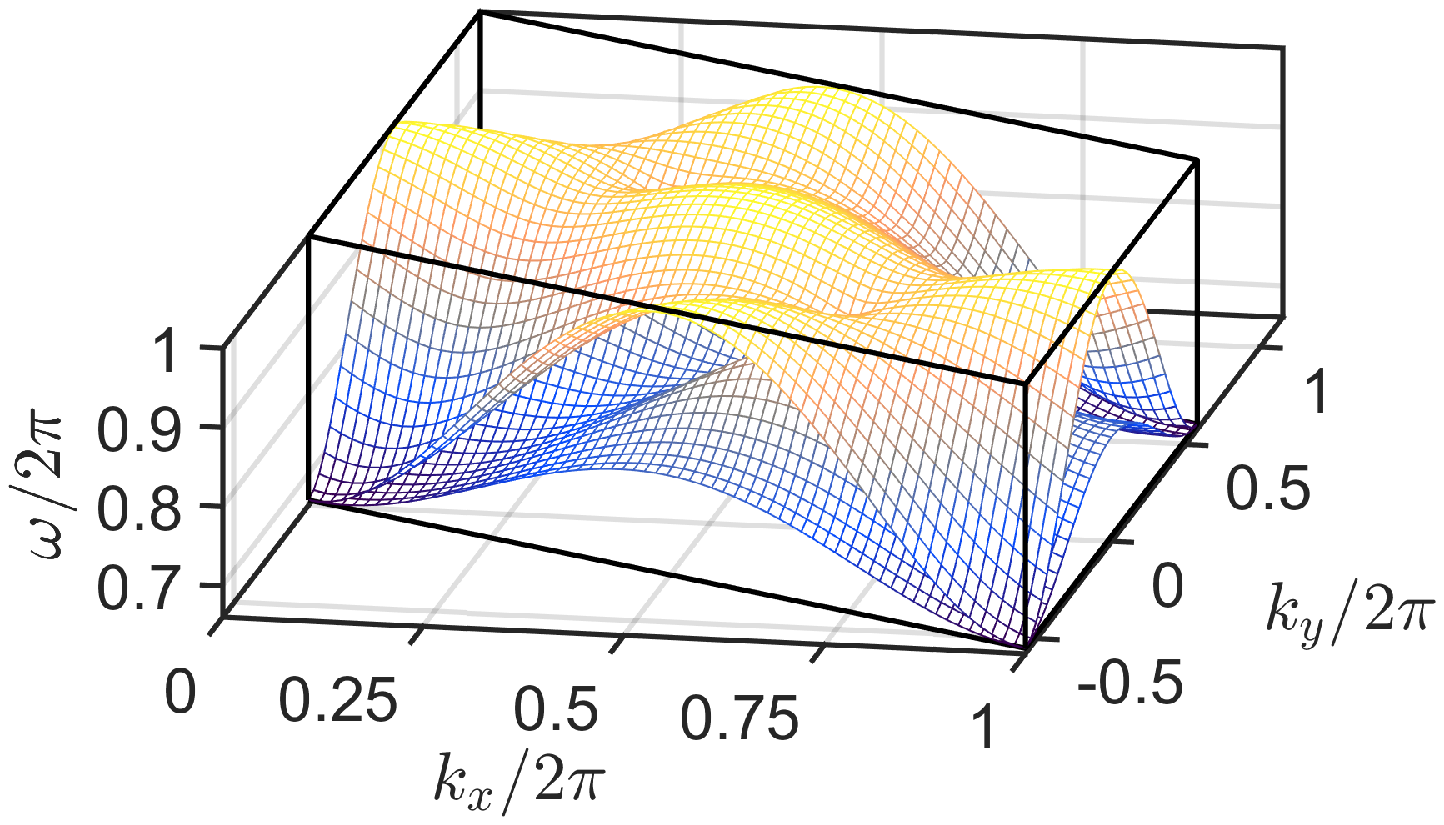}
\includegraphics[width=0.49\textwidth]{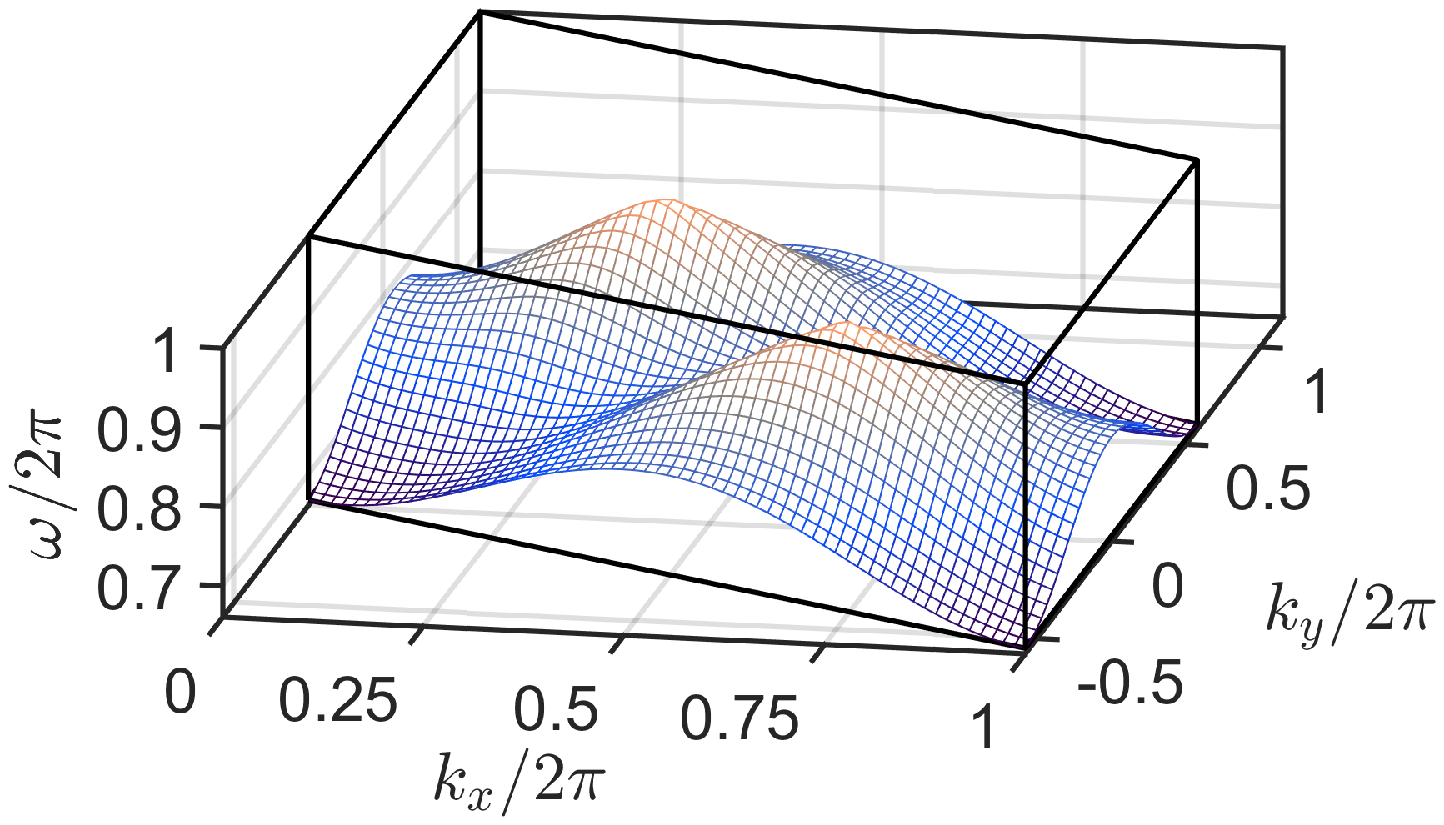}
\end{center}
\caption{Plot of the phonon bands within a reciprocal unit cell. ({\bf Left}): Both upper and lower bands. ({\bf Right}): Lower band. Both bands touch at the two maxima of the lower band and at the corners of the reciprocal cell. Note that paths between corners of the reciprocal cell are equivalent to the path along the short diagonal, but quite different from the long diagonal one.
 }
\label{fig_phonon3D}
\end{figure}
\subsection{Phonons and the 1st Brillouin zone}
\label{subsec:1stBrillouin}
In this subsection we will work in the $\q$ wavevector space for convenience to avoid a factor of 2$\pi$ everywhere.
Primitive vectors of the reciprocal lattice expanded by $\bq1$  and $\bq2$  represent planes of atoms separated by a distance
$1/|\bq{i}|=\sqrt{3}/2$, and perpendicular to them, i.e., at  $ -60^\circ$ with the $x$-axis and horizontal, respectively. Corresponding plane waves propagate in the directions of $\bq{i}$. Plane waves that propagate parallel to the $x$-axis do so in the direction of the smallest reciprocal wavevector
$2\e1=2[1,0]=2\bq1+\bq2$ defining vertical planes separated by a distance $1/2$. Therefore, they have wavevectors with $q_2=2q_1$. The 1st Brillouin zone is a regular hexagon centered at $\Gamma=(0,0)$  with sides 2/3, aphotem $1/\sqrt{3}$, maximal radius 2/3,  with two horizontal sides parallel to the horizontal axis cutting the vertical axis at $1/\sqrt 3$ and two vertices at the $x$-axis at $\pm 2/3$.  Important critical points are vertices $M=(\pm 2/3,0)$ and middle points of the two horizontal sides $K=(0,\pm 1/\sqrt{3})$ and their equivalents through several $\pm 60^\circ$ rotations.

It is convenient to plot the wavevectors in the direction  $[1,0]$ with components along that direction. However, $(1,0)$ is outside the 1st Brillouin zone because planes that bisect $\bq1$ and $\bq2$ cut at vertex $K=(2/3,0)$. The wavevector $[1,0]$ is equivalent to the middle-side point $M=(0,1/\sqrt{3})$, because
$[1,0]-\bq1=[0,1/\sqrt{3}]$, but the meaning of the wavevectors are not so clear. By a $2\times 60^\circ$ rotations M is also equivalent to  $\bq1/2$.

Plots can be done in the Cartesian coordinates both in the direct and reciprocal space because $\e1=[1, 0]$ and $\e2=[0, 1]$ form its own reciprocal basis. We write them as $\r =x\e1+y\e2$ or in the direct lattice coordinates $\r =r_1\a1+r_2\a2$. Similarly, wavevectors can be written as
$\k=k_x\e1+k_y\e2=q_1\bq1+q_2\bq2$. Given the values of $\bq1$ and $\bq2$ above, it is easy to obtain $k_x=2\pi q_1$ and $k_y=2\pi(q_2-\tfrac{1}{2}q_1)\tfrac{2}{\sqrt{3}}$.

\begin{figure}[t]
\begin{center}
\includegraphics[width=0.49\textwidth]{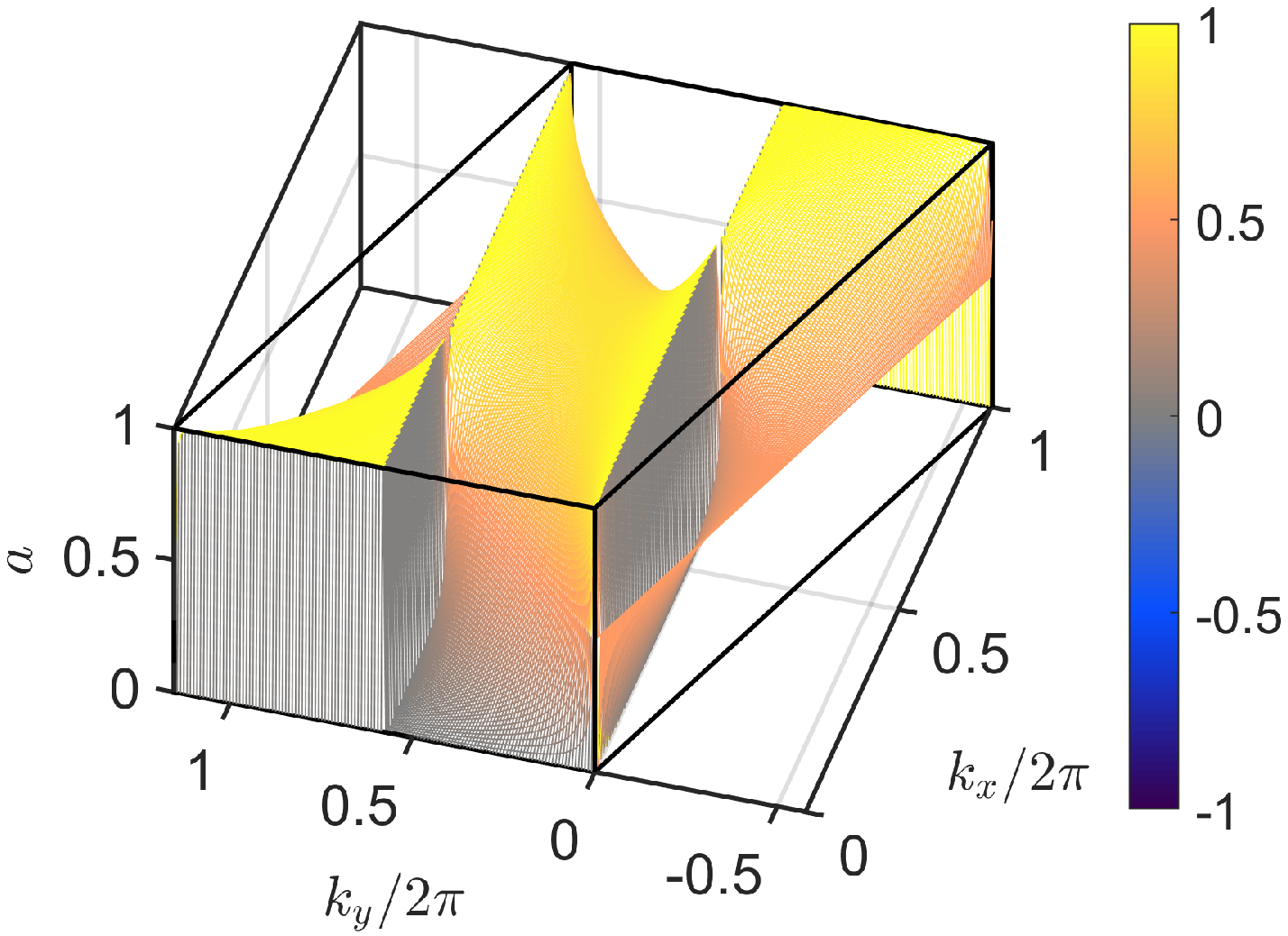}
\includegraphics[width=0.49\textwidth]{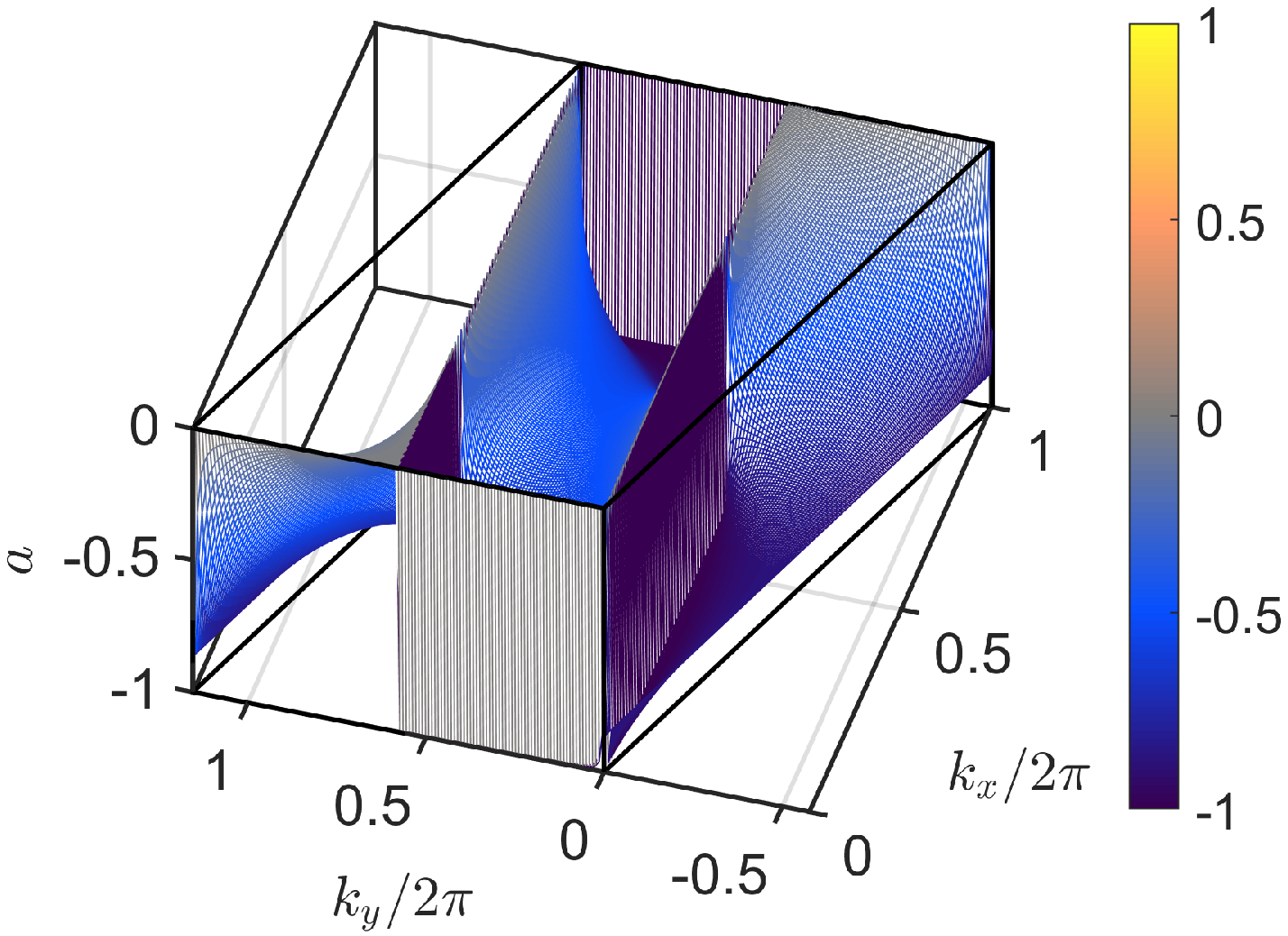}
\end{center}
\caption{Plot of the polarization parameter $a$ with positive sign if sign($a$)=sign($b$) and negative otherwise.
({\bf Left}): Same sign  polarization surface. ({\bf Right}): Opposite sign polarization surface. Note: $a=\pm 1$ indicates that $u_y=0$ and $a=0$ indicates that $u_x=0$.
 }
\label{fig_polarsurf}
\end{figure}
\section{Frequency and polarization of linear modes}
\label{sec:phonons}
In Ref.\,\cite{bajars-physicad2015} the equations that determine the eigenvectors and eigenvalues were deduced. Here we reproduce
the deduction but also obtain the polarization of the phonons. The phonon dispersion relation is plotted along special paths in Fig.\,\ref{fig_BrillouinGMKG} and the phonon surfaces are displayed in a primitive reciprocal cell in Fig.\,\ref{fig_phonon3D}. The polarization is described in  Fig.\,\ref{fig_BrillouinGMKG} and displayed in Fig.\,\ref{fig_polarsurf}.

Linear modes or phonons are given by solutions (to the linearized equations) of the form:
\begin{equation}\label{eq:SimplWave}
 \B{w} = \B{A} e^{\ii\left(\k\cdot \R - \omega t\right)}.
\end{equation}

The variables are as follows: $\B{w}=[u_x,u_y]^T$ is the vector of displacements in the $x$ and $y$ directions of a particle of the lattice and $\B{A}=[a,b]^T$ is its polarization vector. Without loss of generality we can choose $a$ and $b$ such that $\sqrt{a^2+b^2}=1$ and $a\geq 0$. In principle $a$ and $b$ could be complex, but in fact, in our system, they can always be chosen real. $\k$ is a wavevector  and
$\R =n_1\a1+n_2\a2$ is the equilibrium position
corresponding to a given particle. The angular frequency $\omega$ can be chosen positive without loss of generality because the propagation direction of the wave is given by $\k$.

In terms of these indexes the equation above becomes:
\begin{equation}\label{NotSimplWave}
 \B{w}=\B{A} e^{\ii 2\pi\left(q_1 n_1+q_2 n_2- \omega t\right)}=
 \B{A} e^{\ii\left( k_{x}\frac{l}{2}  + k_{y} \frac{\sqrt{3}m}{2} - \omega t\right)},
\end{equation}
leading to the following equation\,\cite{bajars-physicad2015}:
\begin{align*}
\begin{split}
  \B{0} =& \left(\omega^2 - \kappa - 3 \beta\right)\B{A} \\
  & + 2\beta\cos\left(k_{x}\right)\begin{pmatrix} 1 & 0\\0 &
    0 \end{pmatrix} \B{A} + \beta\cos \left( \tfrac{1}{2}k_{x} \right) \cos
  \left( \tfrac{\sqrt{3}}{2}k_{y} \right)
\begin{pmatrix} 1 & 0\\0 & 3 \end{pmatrix}  \B{A} \\
&- \beta\sin \left( \tfrac{1}{2}k_{x} \right) \sin
\left( \tfrac{\sqrt{3}}{2}k_{y} \right)
\begin{pmatrix} 0 & \sqrt{3} \\ \sqrt{3} & 0 \end{pmatrix} \B{A},
\end{split}
\end{align*}
where $\kappa = 16\pi^2/9 = 17.5460$
is the minimal squared frequency, derived from the linearization of the on-site potential in Eq.~\eqref{eq:OnSiteFunc},
and $\beta=V_{LJ}^{''}(1)=72\epsilon=3.6$.

We obtain the homogeneous system of linear equations:
\begin{eqnarray}
  (\omega^2 -A)a-Db=0\, , \nonumber \\
  -D a+(\omega^2 -B)b=0\, ,
  \label{eq:LinearSystem}
\end{eqnarray}
where
\begin{eqnarray}
A= \kappa + \beta \left(3 - 2\cos\left(k_{x}\right) -\cos\left( \tfrac{1}{2}k_{x} \right) \cos
    \left( \tfrac{\sqrt{3}}{2}k_{y} \right) \right)\, ,\nonumber \\
B= \kappa + \beta \left(3 -\cos\left( \tfrac{1}{2}k_{x} \right) \cos \left( \tfrac{\sqrt{3}}{2}k_{y} \right) \right) \, ,\nonumber \\
D=\sqrt{3}\beta\sin \left( \tfrac{1}{2}k_{x} \right) \sin \left(\tfrac{\sqrt{3}}{2}k_{y} \right)\, .
\label{eq:LinearParameters}
\end{eqnarray}
Note that $A\geq \kappa$ and $B\geq \kappa$ and therefore positive, while the sign of $D$ depends on $k_x$ and $k_y$.

The necessary condition for the existence of nonzero solutions of the system in Eq.\,(\ref{eq:LinearSystem}) is the characteristic equation:
$(\omega^2-A)\times(\omega^2-B)-D^2=0$ or $(\omega^2)^2-(A+B)\omega^2+(A B-D^2)=0$, that is:
\begin{eqnarray}
\omega_{1,2}^2=\tfrac{1}{2}\left(A+B\pm\sqrt{(B-A)^2+4 D^2}\right)\, .
\end{eqnarray}
As $\omega>0$, there are two phonon bands corresponding to the two signs of $\pm$. Let us suppose initially that $D\neq 0$, which also implies that $a\neq 0$. For the frequencies $\omega_{1,2}$, the two equations in Eq.\,(\ref{eq:LinearSystem}) are equivalent and only one is needed, for example, the first one. Let us denote $\alpha=b/a$, then
\begin{eqnarray}
\alpha_{1,2}=\frac{\omega_{1,2}^2-A}{D}=\frac{1}{2D}\left( (B-A)\pm \sqrt{(B-A)^2+4D^2}\right)\,   \quad (D\neq 0)\, .
\end{eqnarray}
The normalized polarization vector becomes $a_{1,2}=1/\sqrt{1+\alpha_{1,2}^2}$ and $b_{1,2}=\alpha_{1,2}/\sqrt{1+\alpha_{1,2}^2}$.

It turns out that $D=0$ is an interesting case. In this case, there are two possibilities, either $\sin(k_x/2)=0$ or $\sin(\sqrt{3}/2 k_y)=0$.
In the first one $\sin(k_x/2)=0$, which implies that $k_x=0$ as $k_x=2\pi m$ is equivalent, then:
\begin{eqnarray}
\mathrm{If}\quad k_x=0, \quad \mathrm{then}&\quad q_1=0\quad \mathrm{and}\quad \tfrac{\sqrt{3}}{2} k_y=2\pi q_2\, ,\hfill\mbox{}\nonumber\\
a=0; b=1;\quad  &\omega_1^2 = B=\kappa + \beta \left(3 - \cos \left( \tfrac{\sqrt{3}}{2}k_{y} \right)\right) \, , \hfill\mbox{}\nonumber\\
a=1; b=0;\quad  &\omega_2^2 = A=\kappa + \beta \left(1 -\cos\left( \tfrac{\sqrt{3}}{2}k_{y} \right) \right)\, .
\end{eqnarray}
 These functions reproduce the curves in Fig.\,\ref{fig_BrillouinGMKG}-right and also identify that the upper curve corresponds to $a=0$, that is, there is no vibration in the $u_x$ coordinate, while the lower curve corresponds to $b=0$, that is, there is no vibration in the $u_y$ coordinate.

For the second possibility, $\sin\left(\tfrac{\sqrt{3}}{2}k_{y}\right)=0$, then $\tfrac{\sqrt{3}}{2}k_y=m\pi$ with $m\in\{0,1\}$ and:
\begin{eqnarray}
k_y=\tfrac{2}{\sqrt{3}}\{0,\pi\}, &\quad q_2=\tfrac{1}{2}q_1+\tfrac{1}{2}\{0,1\}\quad \mathrm{and}\quad k_x=2\pi q_1\, ,\hfill\mbox{}\nonumber\\
a=1; b=0;\quad  &\omega_1^2 = A=\kappa + \beta \left(3 -2 \cos\left(k_{x} \right) -(\pm) \cos \left( \tfrac{1}{2}k_{x} \right) \right)\, ,\nonumber\\
a=0; b=1;\quad  &\omega_2^2 = B=\kappa + \beta \left(3 -(\pm)3\cos\left(\tfrac{1}{2}k_{x}\right) \right)\, , \hfill\mbox{}
\end{eqnarray}
where the $+$ or $-$ in $\pm$ corresponds to $m=0$ and $m=\pi$, respectively.
\begin{figure}[t]
\begin{center}
\includegraphics[width=0.49\textwidth]{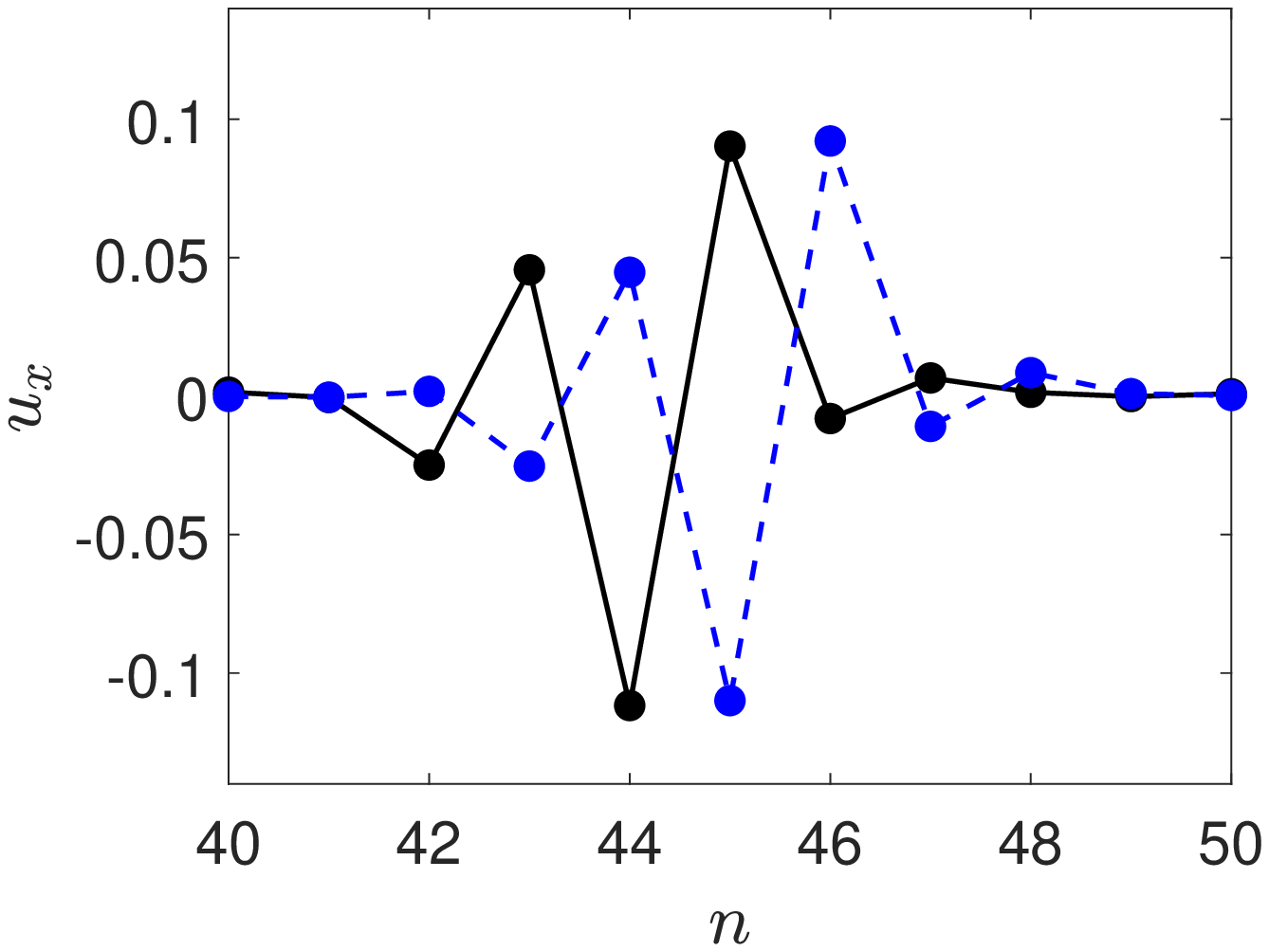}
\includegraphics[width=0.49\textwidth]{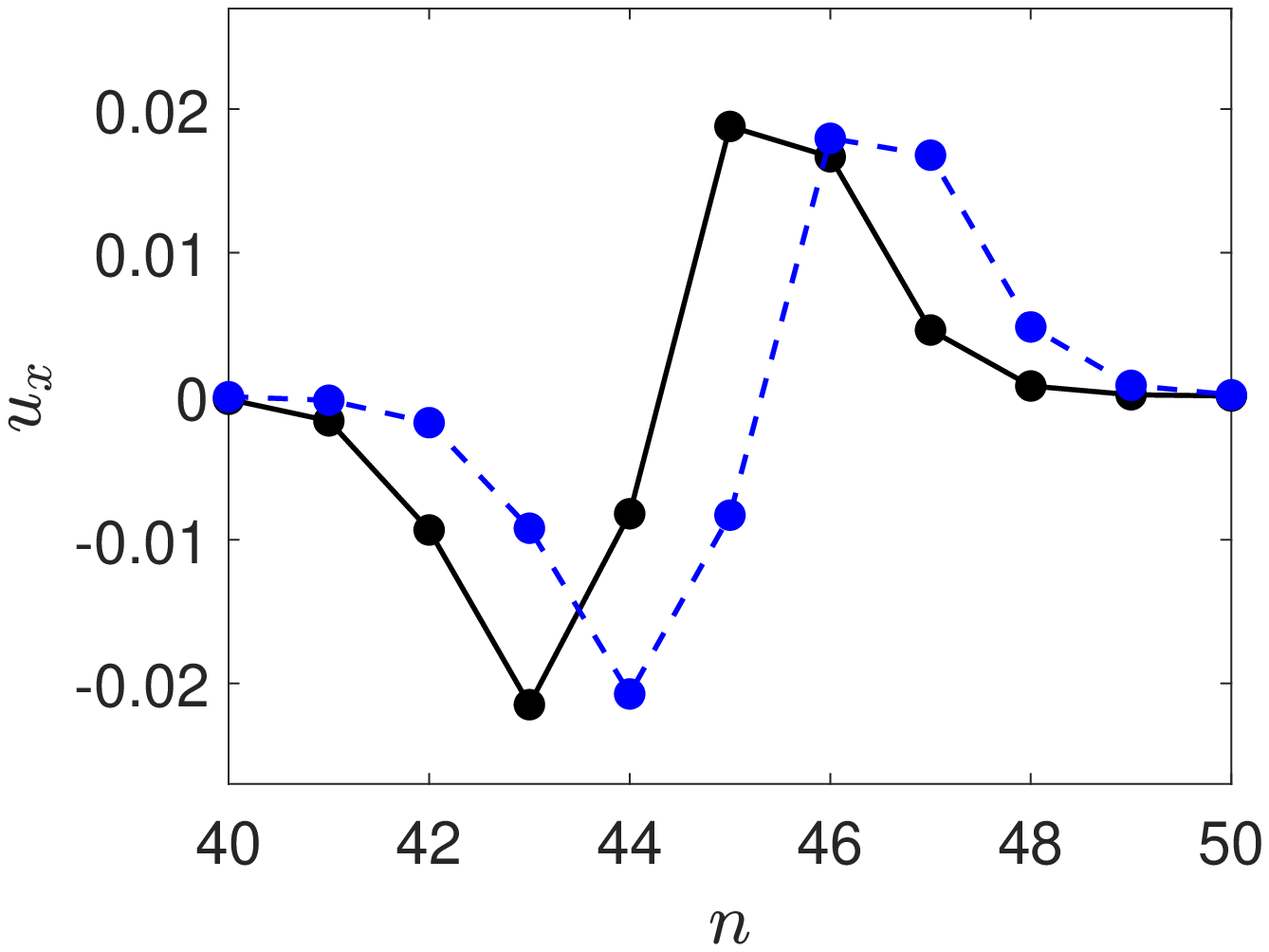}
\end{center}
\caption{{\bf (Left)} Profile of the ILM in Fig.\,\ref{fig_zz_ux} at two times separated by the fundamental time $T_F=s/V_b$, with step $s=1$ showing that the ILM is quasi-exact. See text.
{\bf (Right)} Soliton corresponding to the lower line of the same figure. Data: $\gamma=0.5$.
 }
\label{fig_zz_uxnn_soliton1}
\end{figure}

The curves corresponding to $(\pm)=+1$ correspond to the curves in Fig.\,\ref{fig_BrillouinGMKG}-left. The upper one have polarization $a=1$ for $q_1<2/3$ at point $K$ and $a=0$ for $q_1>2/3$. The lower curves have the opposite polarization. The curves with $(\pm)=-1$ can also be observed in the phonons that are in the phonon band depicted in Fig.\,\ref{fig_wqgamma0-5} right. The phonon surfaces can be seen in Fig.\,\ref{fig_phonon3D}.

The breathers described in this work  derive from the first maximum at point K in the 1st Brillouin zone and, therefore, the central mode of the components of a breather propagating in the $k_x$ direction is a mode with polarization $b=u_y=0$, that is, it has no perturbation in the $y$ direction. This is a very favorable circumstance because, in this way, it tends not to perturb the adjacent chains. Note also that for $D\neq0$, $D$ and therefore $\alpha$ changes sign with the sign of $k_y$, therefore, the sum of plane waves with equal amplitude, the same $k_x$ and opposite $k_y$ and  $b$ will cancel out  leading to polarization $u_y=0$ for the breather, that is, with vibration only in the $x$ directions. This is what happens with the breathers observed in our system.
\begin{figure}[t]
\begin{center}
\includegraphics[width=0.7\textwidth]{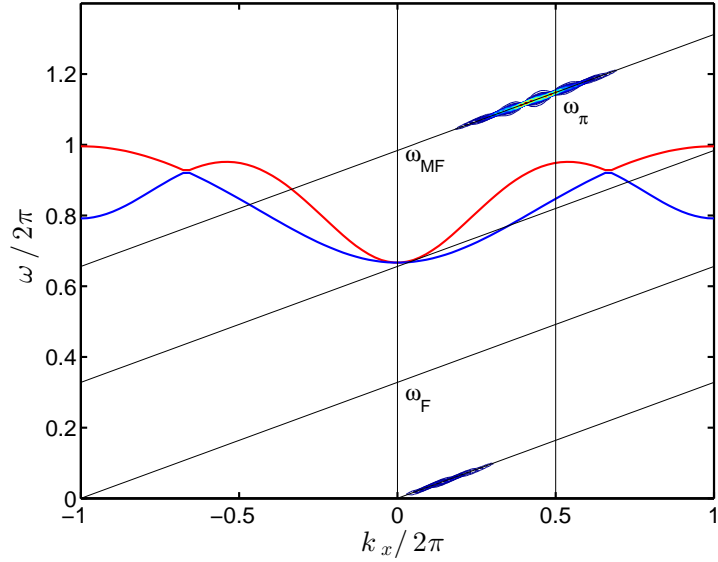}
\end{center}
\caption{XTFT of a long lived ILM (shown in Fig.\,\ref{fig_zz_uxnn_soliton1}) propagating along a close-packed chain corresponding to the primitive vector $\a1$ of the hexagonal lattice. The ILM is composed of a breather along the breather line and a soliton along the soliton line. The breather line is defined by two  frequencies corresponding to $k_x=0$ and $k_x=\pi$, $\omega_{MF}$ and $\omega_\pi$ (or rest frequency), respectively. The breather velocity is $V_b=(\omega_{MF}-\omega_\pi)/\pi$. The breather line is given by $\omega_L=\omega_{MF}+V_b k_x$ and the soliton line by $\omega_L=V_b k_x$. Breather frequencies in the moving frame are just one $\omega_L-V_b k_x=\omega_{MF}$, which, therefore is the breather frequency in the moving frame. There are no intensities at the intersections of the breather line and the dispersion relation, indicating the absence of wings or their small size. Also shown the resonant lines $\omega_L=m\omega_F+V_b k_x$ for $m$ integer, $m=3$ for the breather line.
Data: $\gamma=0.5$. 
 }
\label{fig_zz_ux}
\end{figure}

\section{Basic theory}
\label{sec:basictheory}
In this section we review the theory of moving nonlinear excitations developed in Ref.\,\cite{archilla2019} and extend it to two dimensions. It is presented in a heuristic way together with results for traveling waves in our system. The ideas developed
 here use the fact that a periodic function in the supercell described in Sect.\,\ref{sec:hexagonal} can be expressed as a sum of plane waves with wavevectors in the corresponding momentum space. The amplitudes of these plane waves are obtained by the Fourier Transform (FT). We will specify often the variables that are considered, as, for example, XYTFT indicates the Fourier transform on the two spatial variables and time.

\subsection{Exact traveling waves in one dimension}
\label{subsec:theory1D}
The concepts explained below are illustrated in Figs.\,\ref{fig_zz_uxnn_soliton1} and \ref{fig_zz_ux}, where the traveling wave along the central close-packed chain is represented in the real and frequency-momentum spaces, respectively.
\begin{description}
\item[Traveling wave]\hfill\\
It is represented by $u(n,t)=f(n-V_b t,\omega_{MF} t)$, being $2\pi$ periodic in the second argument. If it is partially localized in the first argument, it becomes a traveling localized wave. The frequency $\omega_{MF}$ is the frequency in the moving frame, where it becomes the only frequency of a stationary profile.
\item[Solitons, kinks and breathers]\hfill\\
 If $\omega_{MF}=0$ and if $f(\pm\infty,0)=0$,  then $u$ represents a soliton. If $\omega_{MF}=0$ and $f(\cdot,0)$ is only zero at $+\infty$ or $-\infty$ and a constant value at the other infinity, $u$ represents a kink. If $\omega_{MF}\neq 0$ and $f$ is localized in the first argument, $u$  represents a breather. The profiles of a breather and a soliton  are represented in Fig.\ref{fig_zz_uxnn_soliton1}.
\item[Exact traveling wave]\hfill\\
If there is a minimal time $T_F$ and integer $s$ such that $u(n+s,t+T_F)=u(n,t)$, then $u$ is an exact traveling wave. A quasi-exact traveling wave can be seen in Fig.\,\ref{fig_zz_uxnn_soliton1}
\item[Fundamental time $T_F$ and step $s$]\hfill\\
The parameters $T_F$ and $s$ are denominated fundamental time and step, respectively. They are related by the velocity of propagation $V_b=s/T_F$.
\item[Fundamental frequency]\hfill\\
It is defined as $\omega_F=2\pi/T_F$. The condition of an exact traveling wave applied to $u(n,t)$, that is, $u(n+s,t+T_F)=u(n,t)$, implies that
$\omega_{MF}=m\omega_F$, where $m$ is an integer. More generally, an exact traveling wave is given by a sum of functions $u$ with moving-frame frequencies that are integer multiple of $\omega_F$.
\item[Resonant plane waves and resonant lines]\hfill\\
   For a given step $s$ and $T_F$ and therefore velocity $V_b=s/T_F$, resonant plane waves are plane waves $u=\exp(\ii[kn-\omega_L t])$ that are exact with the same step $s$ and $T_F$.  Therefore, they can be cast in the form $u=\exp(\ii k(n-V_b t))\exp(-\ii m\omega_F)$. The moving frame frequencies  $m\omega_F$ are related  to the laboratory frequencies $\omega_L$ by:
   \begin{equation}
   \omega_L=m\omega_F+V_bk\, .
    \label{eq:resonatlines}
   \end{equation}
    In the $\omega-k$ representation the above equation represents straight lines called {\em resonant lines} that cut the vertical axis $k=0$ at the moving frame frequencies $\omega_{MF}$ at integer multiples of $\omega_F$. They can be seen in Fig.\,\ref{fig_zz_ux}.

\begin{figure}[t]
\begin{center}
\includegraphics[width=0.49\textwidth]{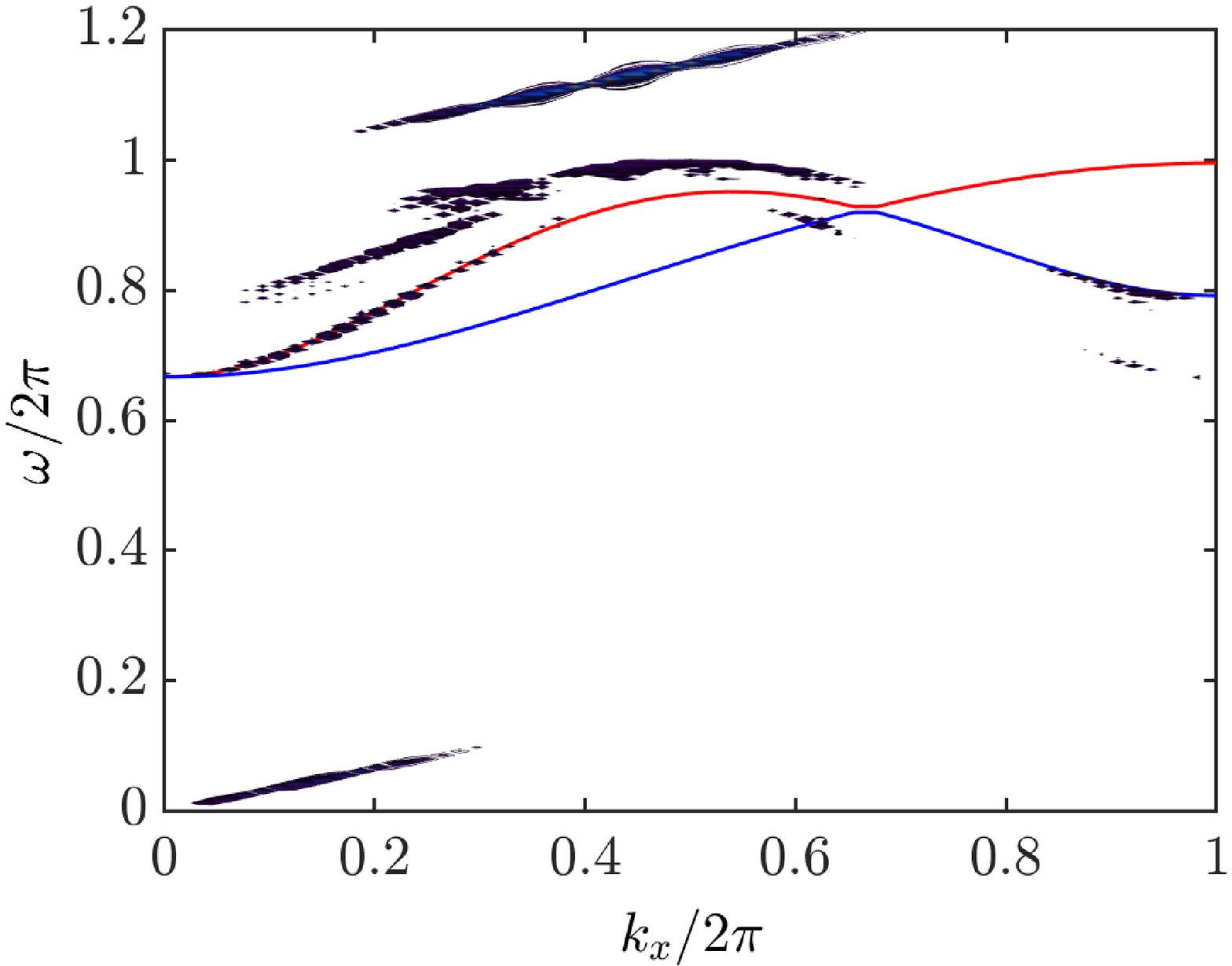}
\includegraphics[width=0.49\textwidth]{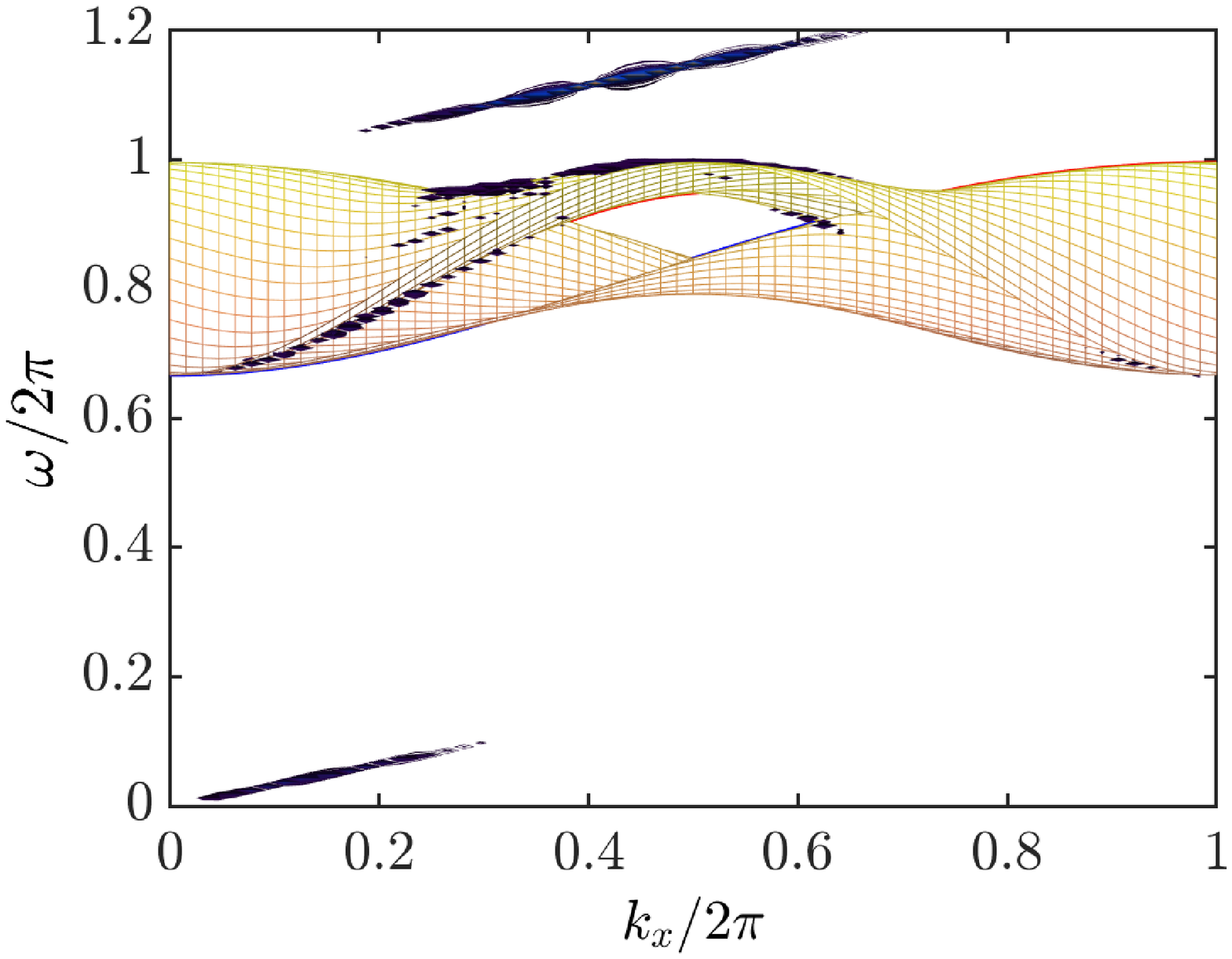}
\end{center}
\caption{{\bf (Left)} Projection of the XYTFT of a long lived  ILM propagating along a close-packed chain of the hexagonal lattice
 corresponding to the $x$-axis. (Blue-Dashed line): dispersion relation along that direction $k_y=0$. (Red and blue lines). It can be seen that there are multiple intensities outside the breather/soliton lines and the dispersion relation.
{\bf (Right)} Same plot but with the projection of the whole phonon bands, suggesting that the intensities are  phonons in other directions, as confirmed in Fig.\,\ref{fig_threeplanes}.
See text. Data: $\gamma=0.5$.
 }
\label{fig_wqgamma0-5}
\end{figure}

\item[Breather line]\hfill\\
    An exact traveling solution is a sum of resonant plane waves. Therefore, in its XT Fourier transform (XTFT) most of the intensities lie on a resonant line, called the \emph{breather line} or on a few of them. If there is only one, then the breather line cuts the axis $k=0$ at the moving frame frequency $\omega_{MF}=m_b\omega_F$ and we recover a single frequency of a breather as in the common stationary case. See Fig.\,\ref{fig_zz_ux}.
\item[Wings] \hfill\\
    Often, there are intensities of the XTFT of an exact traveling wave at the crossing points of a resonant line and the phonon dispersion relation. This means that the traveling waves travel together with one or several resonant phonons. They are called {\em wings}, and should not be confused with tails, that are the diminishing amplitudes from the core of a traveling wave, because the wing amplitudes tend to be constant far from the core of the traveling wave. See Fig.\,\ref{fig_zz_ux}.
\item[Rest or $\pi$ frequency] \hfill\\
      Breathers in hard potentials typically derive from a maximum of $\omega$ at $k=\pi$. The breather mode at $k=\pi$ does not translate and, therefore, its frequency is called the {\em rest frequency} or $\pi$-frequency. Often, moving breathers are obtained by perturbing a stationary breather with wavenumbers centered at $\pi$ and they are a nonlinear perturbation of the $\pi$ phonon. Its value is $\omega_\pi=(m_b+s/2)\omega_F$, and therefore, if $s$ is odd, it is a semi-integer multiple of $\omega_F$. For breathers derived from a $\omega(k)$ minimum at $k=0$, the rest frequency coincides with the frequency in the moving frame.
        \end{description}
\begin{figure}[t]
\begin{center}
\includegraphics[width=0.7\textwidth]{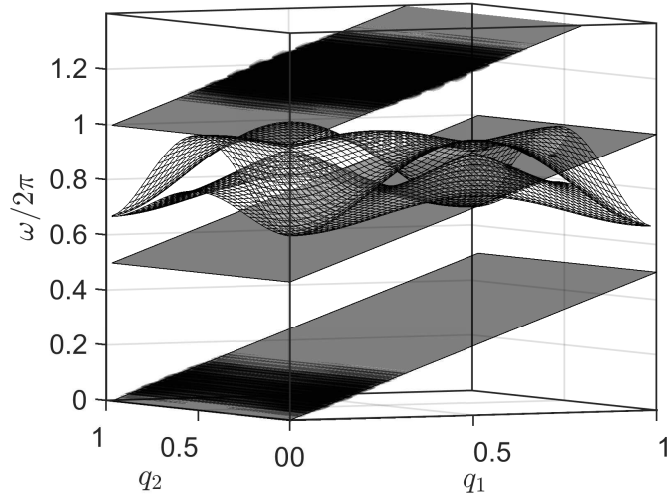}
\end{center}
\caption{Isosurface of the XYTFT of  an exact soliton-breather with $m=2$, together with the resonant planes and the phonon surfaces.
See text. 
Note that the coordinates are the reduced wavenumbers such that $\k=q_1\b1+q_2\b2$.
 }
\label{fig_threeplanes}
\end{figure}

\subsection{Exact traveling waves in two dimensions}
\label{subsec:theory2D}
\begin{figure}[t]
\begin{center}
\includegraphics[width=0.7\textwidth]{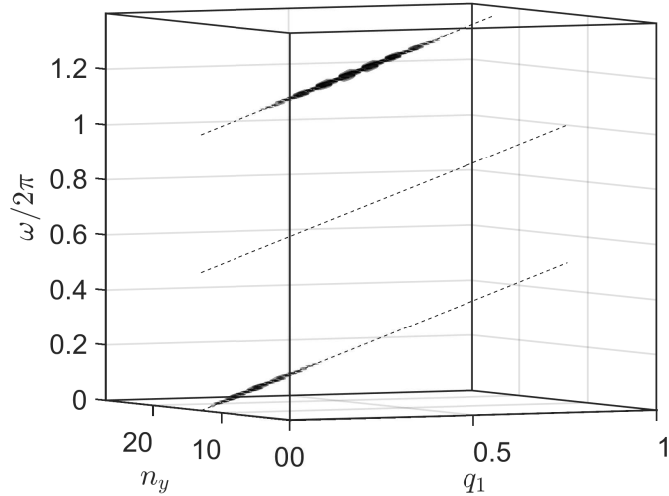}
\end{center}
\caption{Mixed coordinates $n_y$ and XTFT of an exact soliton-breather showing the extreme localization in the $n_y$ direction.
See text. 
Reduced wavenumber $q_1=k_x/2\pi$.
 }
\label{fig_fftmixed}
\end{figure}

Most of the theory in two dimensions is similar to one dimension, with some obvious changes. The variable $u$ may have two components $(u_x,u_y)$ and depends on two indexes in the plane, i.e., $u=u(n_1,n_2,t)$, with $n_1$ and $n_2$ indexes in a sublattice of the Bravais lattice. These concept are illustrated in Fig.\,\ref{fig_wqgamma0-5} as a projection on the $\omega-k_x$ plane and in three dimensions in Fig.\,\ref{fig_threeplanes}.
\begin{description}
\item[Exact traveling waves in 2D]\hfill\\
If we suppose that the direction of a propagation is along $n_1$, then a traveling wave is represented by $u=f(n_1-V_b t,n_2,\omega_{MF}t)$, being $f$ localized in the first two variables and $2\pi$ periodic in the third variable. It is exact if
$u(n_1+s,n_2,\omega(t+T_F))=u(n_1,n_2,t)$ for some step $s$ and fundamental time $T_F$.
\item[Resonant planes and breather plane]\hfill\\
Resonant lines become {\em resonant planes} in the $\omega-k$ space. The breather line becomes the {\em breather plane} which is parallel to the $n_2$ direction. The intensities of the $XYTFT$ are within the breather plane(s). If the traveling wave is very much localized along
 a given direction, the plane may be formed from parallel lines perpendicular to it in $\k$-space.
 They can be seen in Fig.\,\ref{fig_threeplanes}. If the Fourier transform is done only on the $n_1$ variables, then the breather lines reappear as can be seen in Fig.\,\,\ref{fig_fftmixed}.
\item[2D wings]\hfill\\
Wings may appear at the intersections of the resonant planes and the phonon surfaces. They are therefore more complex than in 1D, since they tend to involve more wavevectors, but the frequencies in the moving frame are still a discrete small set corresponding to the few resonant planes that cut the phonon surfaces.  See Fig.\,\ref{fig_threeplanes}.
\end{description}

\subsection{Travelling localized waves in three dimensions and in physical crystals}
\label{subsec:theory3D}
There is no special difficulty to extend the theory to three dimensions, although the visualization and understanding becomes problematic as the frequency-momentum space has four dimensions. Resonant planes becomes hyperplanes and so on. Visualization in 3D will be projections of the 4D $\omega-k$ space.

Then, we can think at how quasi-exact travelling localized waves might appear in actual spectra of physical crystals. Most likely, they will be not so much localized as in simulations, therefore they will be some platelet within a plane which is close to tangent to some slope around a maximum or minimum of the phonon bands. There will be not a single travelling wave but the whole spectrum of them according to their probability of formation. With different velocities and directions they will form a thick truncated cone half filling a valley or a mountain-top in the phonon hypersurfaces. Similar structures have been observed in spectra of materials as SePb at high temperature\,\cite{manley2019} (See Fig. 2 in that reference), which seems promising.

\section{Quasi-exact soliton-breather in the central row. Finding $T_F$.}
\label{sec:quasiexact}
We explore different simulations for good traveling solutions that can be the seed to obtain exact traveling solution as explained in the next section. In general, the procedure is to find a long-lived one, add dissipation at the borders of the simulation cell parallel to the close-packed line where the initial perturbation has been provided and thereafter let it propagate some time without dissipation. In our study we initiate traveling solutions with the velocity pattern in Eq.~\eqref{eq:pattern} and different values of $\gamma$. Such excitation produces only small amount of phonon background.

\begin{figure}[t]
\begin{center}
\includegraphics[width=0.49\textwidth]{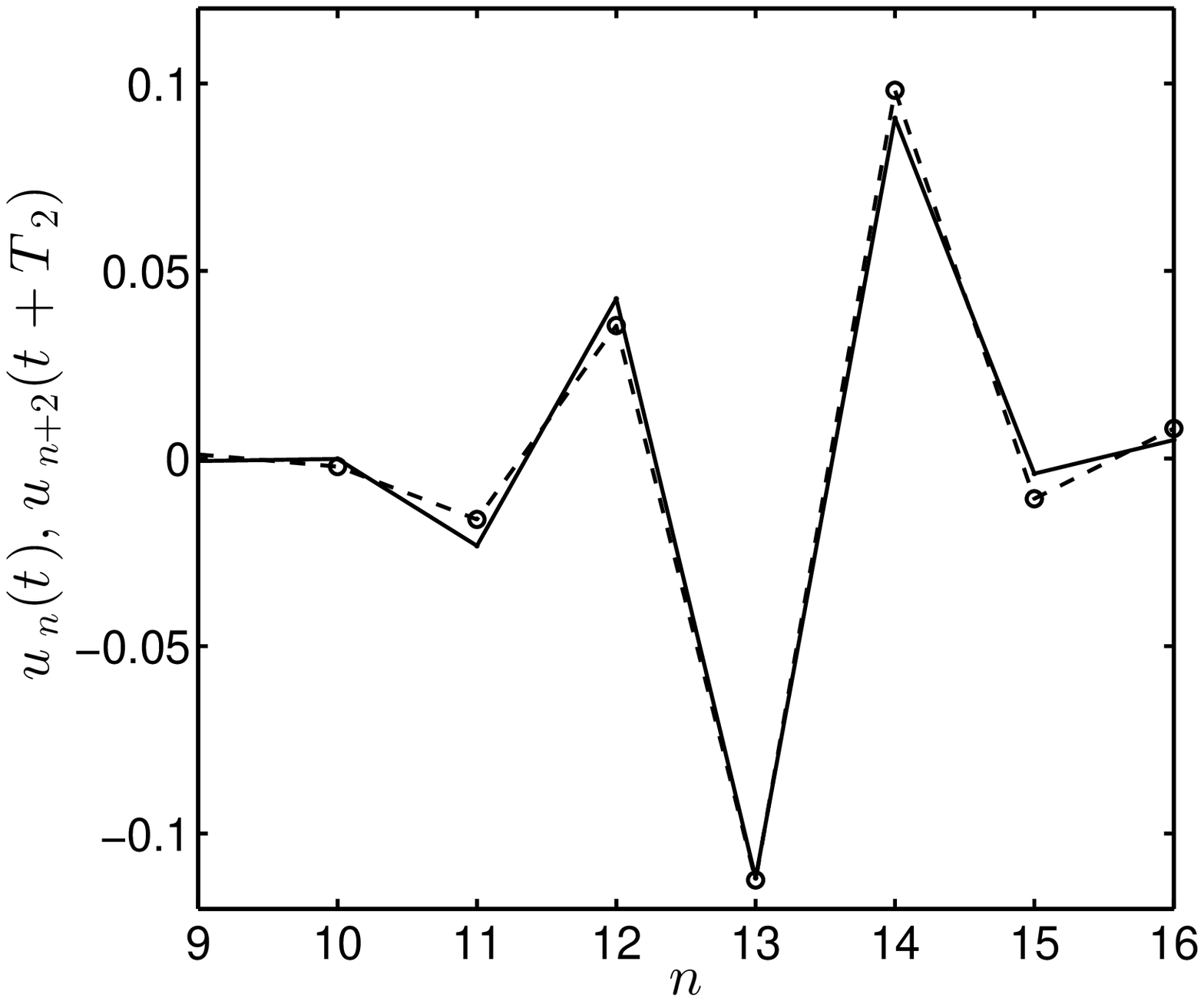}
\includegraphics[width=0.49\textwidth]{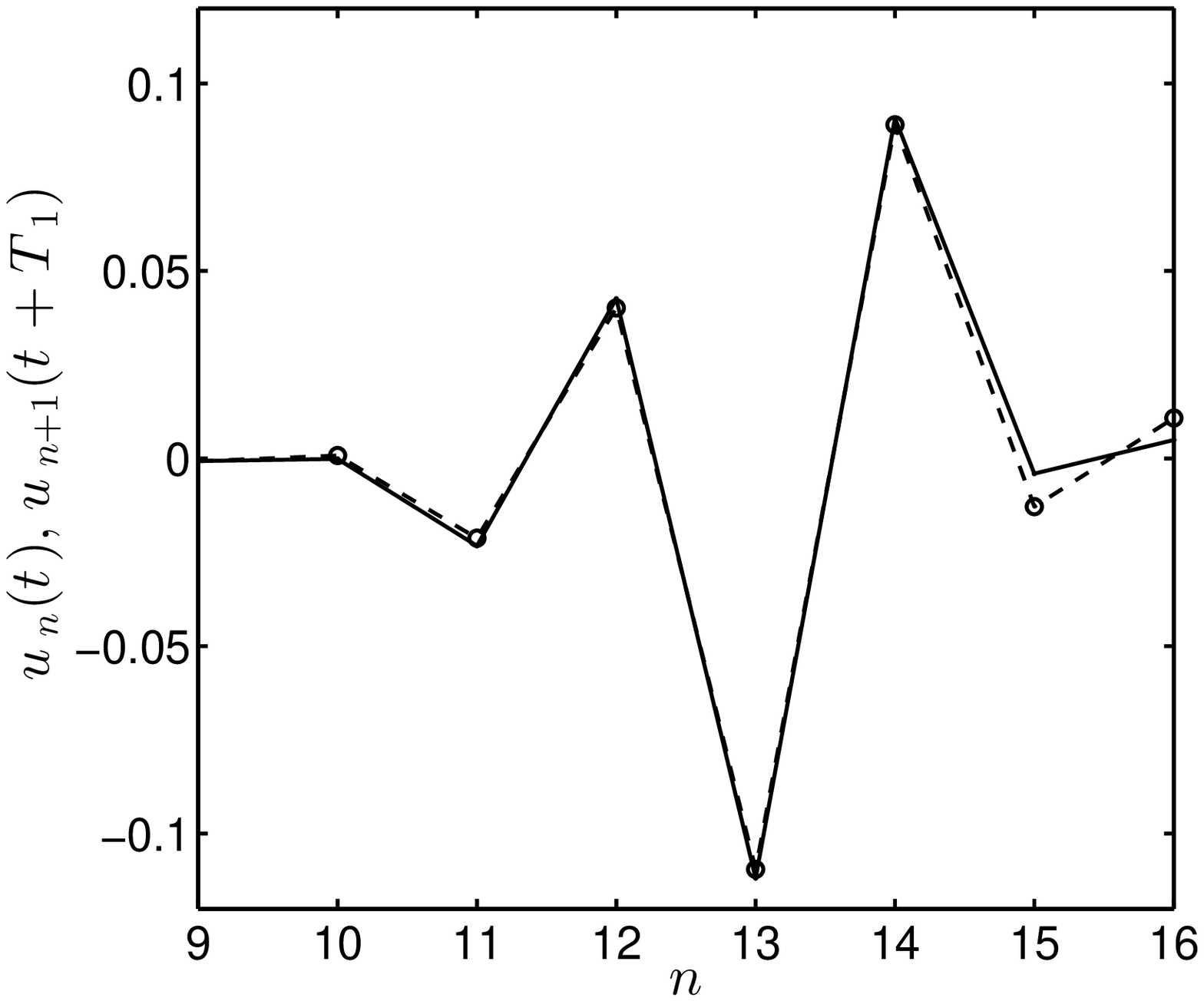}
\end{center}
\caption{ Finding the step $s$ and fundamental time $T_F$. Observation of a 2D breather profiles in its propagation line, $u_n(t)$ (lines)  and its translation a step $s$ and a time $T_s=s/V_b$, that is,  $u_{n+s}(t+T_s)$ (circles). At the left for $s=2$ and at the right for $s=1$. Although the agreement is better for $s=2$, it is clear that $s=1$ is good enough and it is not inverted as would happen if $s=2$. Therefore the closest exact breather has step $s=1$ and fundamental time $T_F\simeq T_1$.
See text. Data: $\gamma=0.5$.
 }
\label{fig_findingTF}
\end{figure}

If we observe the particles in the central line, we can label them with just an index $n\in\{0:N_1-1\}$, i.e., $u_n(t_l)$, with $t_l=l/N_t$, and $l\in\{0:N_t-1\}$.  From the breather line in the XTFT we obtain the breather velocity $V_b=\Delta \omega/\Delta k$. An exact solution has the property that $V_b=s/T_F$, with $s$, the step, and $T_F$, the fundamental time. Then $T_F=s/V_b$ is a multiple of the velocity period $1/V_b=1/f_b$. We consider times $T_s=s/V_b$ with $s=1,2,..$ which are candidates to be $T_F$. We find a time $t_0$ for which the breather amplitude is at maximum and compare the functions $u_n(t_0)$ and $u_{n+s}(t_0+t_s)$, being $t_s=\mathrm{round}(T_s/\tau) \tau$, where $\mathrm{round}(x)$ gives the closest integer to $x$ as $t_s$ is the closest integration time to $T_s$.  Usual steps are $s=1$ or $s=2$\,\cite{archilla2019} and we check the overlapping of the functions. Of course, it is possible to automatize the process, but in practice it is not worth it. Figure\,\ref{fig_findingTF} shows a good example. Velocity was observed to be $V_b=0.3147$, therefore, candidates for fundamental time are $T_1=1/V_b=3.1776$ and $T_2=2/V_b=6.1369$, with integration step $\tau=0.04$, $t_1=\mathrm{round}(T_1/\tau)=76\tau=2.9600=3.1600$ and $t_2=\mathrm{round}(T_2/\tau)=152\tau=2.9600=6.3600$. Mathematically, $s=2$ results in a good agreement, but $s=1$ is good enough and if $s=2$ and $T_F=T_2$, then at midtime $T_F$ the coordinates would be inverted, which does not happen. Therefore, $s=1$ and $T_F$ is close to $T_1$.
   The other adjustment for $T_F$ is that the breather line for $k=0$ is the breather frequency in the moving frame is multiple of the fundamental frequency $\omega_{MF}=m_b \omega_F=m_b\, 2\pi/T_F$. In this way a small adjustment of $T_F$ is possible. Note that the agreement is not perfect, first, due to the fact that the breather is not exact and, second, due the finite time sampling interval $\tau$.

\section{Exact traveling solutions and the Newton method}
\label{sec:newton}
Breathers obtained by providing some recoil velocity are interesting as this is a likely mechanism from $\beta$ decay, as in $^{40}$K, or after the impact of incoming radiation. As they have long lives, they are quite good solutions, however, they are only an approximation to an exact solution. Exact solutions are interesting as they are amenable to mathematical and numerical methods and their lives are theoretically infinite. The generic solutions are breathers accompanied by an extended wing, that is, pterobreathers\,\cite{archilla2019}. In the following, for completeness, we recall some concepts and the method used to obtain exact pterobreathers.

An exact traveling solution is a solution that repeats itself after some time $T_F$, the fundamental time, displaced a lattice step ${\bf s}=[s_1,s_2]$,  where  $s_1$ and $s_2$  ar integers. Let ${\bf n}=[n_1,n_2]$, where $n_1$ and $n_2$ are integers that represent a lattice site $n_1 \a1+n_2 \a2$, and ${\bf U_n}= \{ {\bf u_n }(t), {\bf v_n}(t)\}$ represents the positions and velocities of the particles of the system at site $\bf n$.

Let us define the maps $\hat L[{\bf m}]$, with ${\bf m}=[m_1,m_2]$, and $\hat T[T]$ as the operators of translation in the lattice and time, respectively:
\begin{equation}
\hat L[{\bf m}]:\quad \{{\bf U_n}(t)\}\rightarrow \{{\bf U_{n+m}}(t)\} ;\quad
\hat T[T]:\quad \{{\bf U_n}(t)\}\rightarrow \{{\bf U_{n}}(t+T)\}\, .
\label{eq:transformsLT}
\end{equation}
The action of the operator $\hat T$ is obtained by integrating the equations of dynamics of the system.

The composite translation map in the lattice and time $\hat S[{\bf m},T]$ becomes:
\begin{equation}
\hat S[{\bf m},T]:\quad \{{\bf U_n}(t)\}\rightarrow\{{\bf U_{n+m}}(t+T)\} \,.
\label{eq:transformS}
\end{equation}
Therefore an exact traveling solution with fundamental time $T_F$ and step $\bf s$ is given by the equation:
\begin{equation}
\hat S[{\bf s},T_F] \{{\bf U_n}(0)\}=\{{\bf U_n}(0)\} \,; \quad \textrm{or}\quad \{{\bf U_{n+s}}(T_F)\}=\{{\bf U_n}(0)\}\, .
\label{eq:exactS}
\end{equation}
The election of $t=0$ is irrelevant as the Hamiltonian and dynamical equations are time invariant.
Let as define the function $\bf f$:
\begin{equation}
{\bf f}({\bf U})= \{{\bf U_{n+s}}(T_F)\}-\{ {\bf U_n}(0)\} \, .
\label{eq:fuctionf}
\end{equation}
The exact solution is obtained by the Newton method. Let as suppose that ${\bf U}^0$ is an approximate solution, a seed, that is, ${\bf f}({\bf U}^0)\neq 0$ but small. The approximate  ${\bf U}^0$  can be identified by its initial positions and velocities or by other means as the Fourier coefficients, but this does not affect what follows.
\begin{figure}[t]
\begin{center}
\includegraphics[width=0.7\textwidth]{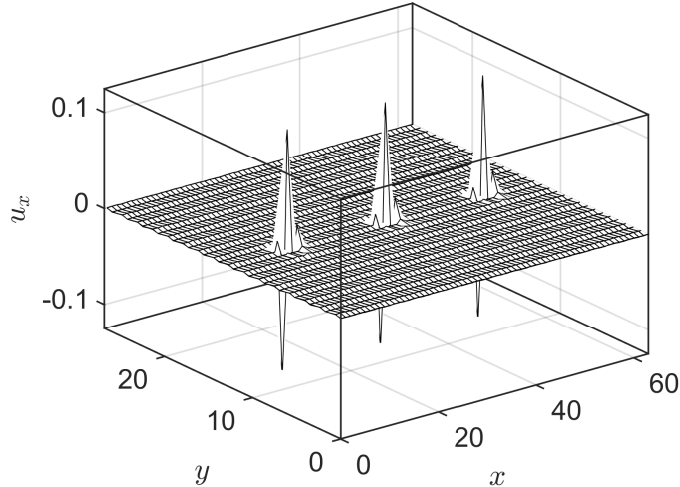}
\end{center}
\caption{Displacements $u_x$  of an exact soliton-breather at three different times separated by $20 T_F$. The localization and exactness can be appreciated.
 }
\label{fig_displ_ux}
\end{figure}

If ${\bf U}={\bf U}^0+\delta {\bf U}$ is an exact solution close to ${\bf U}^0$, i.e., ${\bf f}({\bf U})={\bf f}({\bf  U}+\delta {\bf  U})=0$:
\begin{equation}
{\bf f}({\bf  U})= {{\bf f}}({\bf U}^0)+\nabla {\bf f}({\bf U}^0)\delta {\bf U} \,= 0 , \quad \textrm{and} \quad \delta {\bf U} =- [\nabla {\bf f}({\bf U}^0)]^{-1}{{\bf f}}({\bf U}^0)\, ,
\label{eq:exactU0}
\end{equation}
where $\nabla {\bf f}$ is the Jacobian of ${\bf f}$ calculated at the approximate solution ${\bf U}^0$.

The new initial variables become ${\bf U}={\bf U}^0+\delta {\bf U}$ and they will be closer to an exact traveling solution and becomes a new seed. Repetition of the operation above until the desired accuracy is achieved leads to an exact solution.

Note that it is fundamental to have a good seed and an accurate guess of the step and fundamental time. We are interested, in principle, in traveling wave solutions along close-packed lines, which are most likely to occur in a mathematical and physical system. As all close-packed lines are equivalent, the simplest one is a line parallel to $\a1$ and the step becomes $\mathbf{s}=[s_1,0]$. For simplicity, we will refer often to the step just as the scalar  $s\equiv s_1$. Typical values of $s$ are very low integer numbers 1,2,...

\section{Exact soliton-breathers}
\label{sec:results}
With the velocity pattern in Eq.~\eqref{eq:pattern} we find different quasi-exact traveling waves composed of a breather and a soliton. The soliton is always present and can be reconstructed by filtering out the frequencies above the soliton line and performing the inverse Fourier transform. Figure~\ref{fig_zz_uxnn_soliton1} shows both profiles of the soliton-breather and the soliton. The soliton has a ``S''-shape, a compression followed by a decompression. This is coherent with the physical mechanism for producing propagating lattice excitations, i.e., the impact of a swift particle or the recoil from a decay event as the $\beta$-decay of $^{40}$K. However  similar methods did not produce a soliton-breather but a pure breather in the related model by Archilla et al.\,\cite{archilla2019}. The soliton-breather has a length of 5-6 particles.

\begin{figure}[t]
\begin{center}
\includegraphics[width=0.7\textwidth]{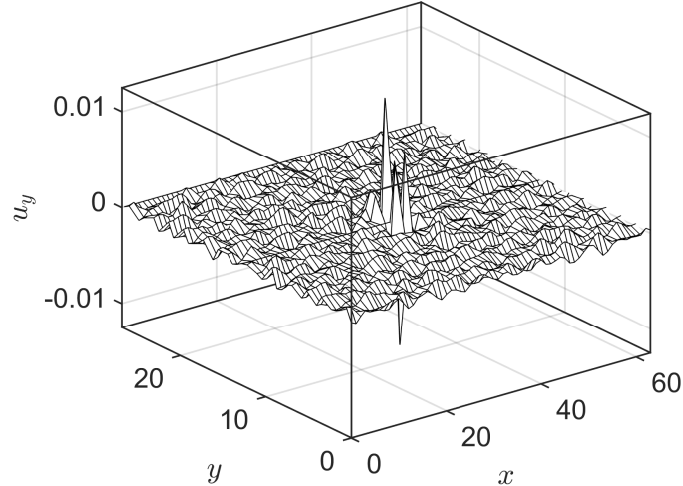}
\end{center}
\caption{Displacements $u_y$ of an exact soliton-breather. Note the smaller value of the amplitude with respect to $u_x$ shown in Fig.\,\ref{fig_displ_ux}.
 }
 \label{fig_displ_uy}
\end{figure}

Many solutions have a long life, specially, if the initial phonons are absorbed by the borders by computational means. These traveling waves are quasi-exact, meaning that the properties of exact traveling breathers hold very well, although not exactly. Using them as a seed and through the Newton method described above, it is possible to obtain exact traveling waves also composed of a breather and soliton and with small amplitude wings.  In Fig.\,\ref{fig_displ_ux} the plot of the particle coordinates $u_x$ is presented for three times separated by $20 T_F$. Both the localization and the exactness can be appreciated. Figure \ref{fig_displ_uy} demonstrates the $u_y$ coordinates. They have similar spectral properties as $u_x$ but their amplitudes are much smaller.

In Table\,\ref{table:exactbreathers} some examples of parameters of exact soliton-breathers are shown. Particular exact traveling wave solutions were obtained on the lattice with dimensions $N_1=32$ and $N_2=16$. Unfortunately, we have been able to find only steps $s=1$ and we presently do not know if this is a characteristic of the system or it shows the need for very different generation methods to produce different seeds.

\begin{table}[h]
 \begin{center}
\begin{tabular}{ |c|c|c|c|c|c|c|c|c|c| }
 \hline
 $\gamma$ & $s$ & $m_b$ & $T_f$ & $V_b$ & $f_{F}$ & $f_{L}(0)$ & $f_{L}(\pi)$ & Abs.~error& Rel.~error\\
 \hline
 0.7 & 1 & 1 & 1.12 & 0.8929 & 0.8929 & 0.8929 & 1.3393 & 8.7441e-14 & 1.2730e-15 \\
 \hline
 0.65 & 1 & 2 & 1.96 & 0.5102 & 0.5102 & 1.0204 & 1.2755 & 1.1302e-13 & 1.7266e-15 \\
 \hline
 0.6 & 1 & 2 & 2.00 & 0.5000 & 0.5000 & 1.0000 & 1.2500 & 9.9292e-14 & 1.6558e-15 \\
 \hline
 0.55 & 1 & 3 & 2.96 & 0.3378 & 0.3378 & 1.0135 & 1.1824 & 1.7740e-13 & 2.1779e-15 \\
 \hline
 0.5 & 1 & 3 & 3.04 & 0.3289 & 0.3289 & 0.9868 & 1.1513 & 1.3878e-13 & 2.1391e-15 \\
 \hline
 0.45 & 1 & 3 & 3.12 & 0.3205 & 0.3205 & 0.9615 & 1.1218 & 1.2790e-13 & 2.0743e-15 \\
 \hline
 0.42 & 1 & 4 & 4.16 & 0.2404 & 0.2404 & 0.9615 & 1.0817 & 1.3585e-13 &  2.4348e-15 \\
 \hline
 0.4 & 1 & 5 & 5.12 & 0.1953 & 0.1953 & 0.9766 & 1.0742 & 1.3310e-13 & 2.8352e-15\\
 \hline
\end{tabular}
\end{center}
\caption{Parameters of exact soliton-breathers. $\gamma$: modulus of initial kick, $s$: step, $m_b=\omega_{MF}/\omega_F$, $T_F$: fundamental time, $V_b$: breather velocity, $f_{F}=\omega_F/2\pi$: fundamental frequency, $f_{L}(0)$, $f_{L}(\pi)$: laboratory frequencies ($\omega_L/2\pi$) at wavenumbers 0 and $\pi$, and absolute and relative errors of $\{u(n+s,t+T_F)-u(n,t)\}$.
} \label{table:exactbreathers}
\end{table}

\section{Conclusions}
We have expanded the theory of exact traveling waves developed for one dimension in a previous publication to two dimensions in a hexagonal lattice, applying it to a model for silicate layers in which some authors have previously found breathers with very long life. The theory is based in the representation in the frequency-momentum space and has allowed the determination of the structure of the propagating waves in the model. They are formed by a breather and a soliton traveling together, that is, a soliton-breather or solbreather. We have described the system in terms of the direct and reciprocal hexagonal lattice, finding the structure of the 1st Brillouin zone including the polarization of the phonon surfaces. The theory describes traveling waves in the $\omega-k$ representation and shows that exact traveling waves lie within parallel planes in that space, each one corresponding to a specific frequency in the moving frame. These frequencies are integer multiples of a minimal one, called the fundamental frequency. One of these planes is the breather plane where the breather lies. The others produce the wings at their intersection with the phonon surfaces. The soliton corresponds to a resonant plane with zero frequency. We have developed a method for obtaining the fundamental time of the traveling waves and used it to obtain exact traveling waves. A variety of them have been found, all of them with small wings and step $s=1$. We are exploring methods to obtain other steps or to find out if they are not possible in our system. The main conclusion is that the theory of exact traveling waves and their spectral representation are powerful means to observe the structure of traveling waves and obtain exact ones. From the physical point of view exact solutions are more likely to appear in physical processes than approximate solutions and their properties can be studied more easily.

\section*{Acknowledgments}
J Baj\={a}rs acknowledges support from PostDocLatvia grant No.1.1.1.2/VIAA/4/20/617. JFR Archilla thanks projects PAIDI 2021/FQM-280 and MICINN PID2019-109175GB-C22.
He also acknowledges a travel grant from VIPPITUS-2020 and the University of Latvia for hospitality.
Both authors aucknowledge Prof. M.E. Manley for useful information and discussions about neutron scattering spectra of some materials.
\appendix
\section{Construction of the reciprocal lattice}
\label{App:reciprocal}
For a hexagonal lattice of unit distance $a$, the direct vectors that generate the lattice are\,\cite{ashcroftmermin}:
\begin{equation}
\a1=a\e1; \quad \a2= \frac{a}{2}\,\e1 +\frac{\sqrt{3}}{2}a\,\e2;  \quad \a3=c \e3 .
\end{equation}
We will not require  $\a3$ but it is convenient to leave it initially.  Any lattice point can be obtained as the Bravais direct lattice:
\begin{equation}
{\R}=n_1 \a1+n_2\a2+n_3\a3\, ,
\end{equation}
with $n_1, n_2, n_3$ integers. The unit cell has a volume $v=\a1\cdot (\a2\times \a3)=(\sqrt{3}/2)a^2 c$.

Many physical properties are described by a function of the direct Bravais lattice, for example the on-site potential as can be seen in Fig.~\ref{fig_Udirect}

The corresponding reciprocal lattice basis can be defined by:
\begin{equation}
\b1=2\pi\frac{\a2\times\a3}{\a1\cdot (\a2\times \a3)}; \quad \b2=2\pi\frac{\a3\times\a1}{\a1\cdot (\a2\times \a3)};  \quad \b3=2\pi\frac{\a1\times\a2}{\a1\cdot (\a2\times \a3)} .
\end{equation}
The resulting basis is:
\begin{equation}
\b1=\frac{2\pi}{a} [1, -\frac{1}{\sqrt{3}},  0] ; \quad \b2=\frac{2\pi}{a} [0, \frac{2}{\sqrt{3}},   0 ];  \quad \b3= 2\pi\frac{2}{\sqrt{3}c} [0 , 0,   1 ],
\end{equation}
with the properties that $|\b1|=|\b2|=2\pi ({2}/{\sqrt{3}a})$, $|\b3|=2\pi({2}/{\sqrt{3}c})$ and $\a{i}\cdot\b{j}=2\pi\delta_{ij}$.

As the third coordinate and dimension is not relevant in our problem, we suppress it, and take $a=1$ as the unit of a distance in the direct space and $1/a$ in the reciprocal space to simplify the notation.

The resulting bases are:
\begin{eqnarray}
\hfill \a1=[1, 0]; \quad \a2= [\frac{1}{2}, \frac{\sqrt{3}}{2}];\quad \hfill\\ \nonumber 
\b1=2\pi[1, -\frac{1}{\sqrt{3}}]=2\pi \frac{2}{\sqrt{3}}[\frac{\sqrt{3}}{2},-\frac{\sqrt{1}}{2} ]; 
\quad \b2=2\pi [0, \frac{2}{\sqrt{3}} ]=2\pi\frac{2}{\sqrt{3}} [0,1].
\end{eqnarray}

They have the property that $\a{i}\cdot\b{j}=2\pi \delta_{ij}$, and therefore if $\K=m_1\b1+m_2\b3$, with $m_1,m_2$ integers, is a vector in the reciprocal lattice then $\K\cdot\R=2\pi (m_1 n_1+ m_2 n_2)=2\pi n $, with $n$ an integer.

\section*{References}
\bibliography{sevilla2021,russell2021}
\end{document}